\documentclass[journal]{IEEEtran}
\usepackage{array}
\usepackage{tabularray}
\usepackage{dirtytalk}
\usepackage{amsfonts} 
\newcolumntype{P}[1]{>{\centering\arraybackslash}p{#1}}
\usepackage{wrapfig}
\usepackage{pgfplots}
\pgfplotsset{compat=1.16}
\usepackage{cite}
\usepackage{multirow}
\usepackage{subcaption}
\usepackage{booktabs,tabularx,graphicx}
\usepackage{amsmath}
\usepackage{stackengine}
\usepackage{balance} 
\usepackage{comment}
\usepackage{float}
\usepackage{algorithm} 
\usepackage{enumitem}
\setlist[itemize]{align=parleft,left=0pt..1em}
\usepackage[font=small,labelfont=bf, textfont=sc]{caption}
\usepackage{lipsum}
\DeclareMathAlphabet\mathzapf       {T1}{pzc} {mb} {it}
\usepackage{rotating}
\usepackage{wasysym}
\usepackage{pifont}  
 \newcommand{\cmark}{\ding{51}}
\newcommand{\xmark}{\ding{55}}
\usepackage{colortbl}
\usepackage{hhline}
\renewcommand{\thesubsection}{\thesection--\arabic{subsection}}

\DeclareUnicodeCharacter{2212}{-}

\begin{document}

\title{{Autoregressive Coefficients based Intelligent Protection of Transmission Lines Connected to Type-3 Wind Farms}}

\author{Pallav~Kumar~Bera,~\IEEEmembership{Member,~IEEE,}
        Vajendra~Kumar,~\IEEEmembership{}
        Samita~Rani~Pani,~\IEEEmembership{}
        Om~P.~Malik,~\IEEEmembership{Life Fellow,~IEEE}
        \vspace{-6mm}

\thanks{Pallav Kumar Bera  is with the School of Engr. \& Applied Sciences, Western Kentucky University, KY  42101, USA (e-mail: pallav.bera@wku.edu).}
\thanks{Vajendra Kumar was with Indian Inst. of Tech., Roorkee 247667, India
(e-mail: kumarvajendra@gmail.com).}
\thanks{Samita Rani Pani is with the EE Dept.,  Kalinga Inst. of Industrial Tech., Bhubaneswar 751024, India (e-mail: samita.panifel@kiit.ac.in).}
\thanks{Om P. Malik is with the ESE Dept., University of Calgary, Calgary, AB
T2N 1N4, Canada (email:maliko@ucalgary.ca).}

\thanks{©2023 IEEE. Personal use of this material is permitted. Permission from IEEE must be obtained for all other uses, in any current or
future media, including reprinting/republishing this material for advertising or promotional purposes, creating new collective works, for
resale or redistribution to servers or lists, or reuse of any copyrighted component of this work in other works.}
\thanks{doi: 10.1109/TPWRD.2023.3321844}
}

\markboth{TO APPEAR IN IEEE TRANSACTIONS ON POWER DELIVERY (PWRD)}
{Bera \MakeLowercase{\textit{et al.}}: Autoregressive Coefficients based Intelligent Protection of Transmission Lines Connected to Type-3 Wind Farms}

\maketitle

\begin{abstract}— 
Protective relays can mal-operate for transmission lines connected to doubly fed induction generator (DFIG) based large capacity wind farms (WFs). The performance of distance relays protecting such lines is investigated and a statistical model based intelligent protection of the area between the grid and the WF is proposed in this article. The suggested method employs an adaptive fuzzy inference system to detect faults based on autoregressive (AR) coefficients of the 3-phase currents selected using  minimum redundancy maximum relevance algorithm. Deep learning networks are used to supervise the detection of faults, their subsequent localization, and classification. The effectiveness of the scheme is evaluated on IEEE 9-bus and IEEE 39-bus systems with varying fault resistances, fault inception times, locations, fault types, wind speeds, and transformer connections. Further, the impact of factors like the presence of type-4 WFs, double circuit lines, WF capacity, grid strength, FACTs devices, reclosing on permanent faults, power swings, fault during power swings, voltage instability, load encroachment, high impedance faults, evolving and cross-country faults, close-in and remote-end faults,  CT saturation, sampling rate, data window size, synchronization error, noise, and semi-supervised learning are considered while validating the proposed scheme. 
The results show the efficacy of the suggested method in dealing with various system conditions and configurations while protecting the transmission lines that are connected to WFs. 

\end{abstract}

\begin{IEEEkeywords}
Intelligent Protection, Fault detection, High Impedance Faults, Power Swings, TCSC, Feature Selection, Fuzzy, InceptionTime, Machine Learning, Wind Farms
\end{IEEEkeywords}

\section{INTRODUCTION}
Increase in power demand, combined with recent research initiatives aimed at increasing the use of wind energy, has resulted in the rapid adoption of wind farms (WFs)\cite{windfarm2015}. Meanwhile, wind-rich regions are usually far from the load centers, necessitating the transportation of generated energy over vast distances through high-voltage transmission lines (TLs). Wind turbine generators (WTGs) are integrated into the grid through power electronic converters. Different fault ride-through (FRT) schemes are employed by various country grid codes which handle over-currents, over-voltages and subsequent instability in the system \cite{gridcode}. The diversity of FRT controls and nonlinear properties of converters alter the fault characteristics of the power system \cite{frt}. Hence, conventional protection based on fundamental frequency phasor measurement faces significant challenges in the presence of large-scale WFs \cite{coventional}. {Also, with the increased integration of WFs, malfunctions of phase selector and directional overcurrent relays have been reported \cite{directional}\cite{cosine22}. Additionaly, the presence of series compensation challenges the operation of distance relays \cite{sauvikstatcom22}.}

Before WTGs are integrated into the main grid, it is critical to study their fault characteristics to better understand the protection requirements.  Fault analysis assists the protection engineers in choosing the appropriate circuit breaker, relay, and protection schemes. The fault currents associated with WFs differ from those associated with conventional synchronous generators due to the power electronics components involved, having low magnitude, increased harmonics, and non-linearities \cite{spearman21}. In contrast to type-1 and type-2 WTGs, where fault characteristics are determined by the system and WTG impedances, type-3 and type-4 WTGs have faults that are complex and depend on the controllers \cite{chen}. The fault current in type-3 WFs has power and non-power frequencies, and fast dc decaying components whereas it contains mainly power and non-power frequencies in the case of type-4 WFs. Since short circuit studies for these WTGs are not universal and are design-specific, an alternate and comprehensive protection mechanism in place of the traditional impedance-type distance relays typically used for primary and backup protection of HV TLs is required.

Dynamic operating conditions in WFs challenge the fixed settings of traditional differential and distance relays.
Recent studies have suggested the use of adaptive distance and differential relay settings to detect faults in TLs connected to WFs. 
Swarm intelligence-based adaptive threshold differential protection is proposed in \cite{PRASAD2020}. An adaptive method to modify the protective settings taking into account the various output power levels of the WF is presented in \cite{saber22}. Using local knowledge of current, voltage, and the number of WF units, an adaptive distance relay strategy is developed to protect the transmission system connected to a WF in \cite{pradhan}. An online adaptive distance relay setting for parallel transmission networks having FACTS with WFs is suggested in \cite{DUBEY2015}.

Pearson correlation coefficient and cosine similarity distance to discriminate internal and external faults in TLs integrated to WFs has been used in \cite{pearson18, cosine, cosine22}. These techniques are ineffective when circuit breakers reclose on permanent faults or the output of WF is poor causing improper operation of proposed pilot protection during normal conditions and internal faults. Signed correlation of the phase currents is used to differentiate faults and other events in \cite{saber22}. However, different aspects, such as high impedance faults, evolving and cross-country faults, stressed conditions, CT saturation, and so forth, are not considered.   

Use of machine learning (ML) techniques to augment the fault detection and classification process has also been proposed by some investigators. Differential protection based on support vector machines to accurately differentiate internal and external faults for intertie zone between a WF and the grid is developed in \cite{rezaei}. Signs of superimposed currents were used for fault detection and section identification whereas dual-time transform and decision tree were used for fault classification in STATCOM compensated TL connecting a WF in \cite{sauvikstatcom22}. For a TCSC compensated line, positive-sequence current-based fault identification and an empirical mode decomposition (EMD) supported random forest classifier are proposed in\cite{sauviktcsc21}. Even though the above work highlights how effective ML can be, they only employ a specific feature for fault analysis. However, it is unlikely that a single feature can encompass all fault characteristics associated with WFs. Before using any learning technique, it is important to assess different features and use a feature selection procedure. 

{In this article, a novel autoregressive (AR) coefficient based intelligent protection scheme that encompasses various power system conditions and considers a range of features for the protection of lines connected to type-3 WFs is proposed.}
A summary of the originality and primary contributions of this study is given below.
 \vspace{-0.5 mm}
\noindent

\begin{itemize}
 \item { An AR coefficient-based intelligent protection to detect, locate, and classify faults is proposed. The approach is tested for different scenarios simulated by considering various parameters that could impact the fault currents. }

   \item { A list of features are evaluated using maximum relevance minimum redundancy (mRMR) algorithm to obtain the top feature which is used to train the fuzzy inference system and various ML algorithms.}
  \item The proposed protection is validated for the presence of type-4 WFs, {multiple type-3 WFs}, weak grids, change in WF units, double circuit lines, FACTS devices, power swings, load encroachment, voltage instability, cross-country and evolving faults, close-in and far-end faults, high impedance faults, auto-reclosing on permanent faults, CT saturation, different sampling frequency, synchronization errors, noise, etc.
   \item { The proposed method does not need extraction of fundamental frequency components; it is sufficient to use instantaneous time domain data of 0.5 cycles. Hence, the proposed protection is free from the influence of off-nominal frequency (42-78 Hz due to variable slip of DFIG).}
  \item { The operation of traditional distance relay in various scenarios is investigated and the performance of the proposed method is compared with existing works.}
     \item The faults and transients dataset\cite{data} of over 40000 cases is uploaded to the IEEE data-port data repository to enable interpretation, replication of  findings, and future use.
\end{itemize}

The remaining portions of the article are organized as follows: Modeling and simulation of faults and non-fault events, and {challenges} to conventional distance relays in the IEEE 9-bus test system are discussed in Sec. II. The AR coefficient-based intelligent protection scheme, feature selection, and deep learning network (InceptionTime) are described in Sec. III. The fuzzy-based decision-making system for fault detection and results of fault detection, location, and phase selection for type-3 WF in IEEE 9-bus system are discussed in Sec. IV.  Sec. V validates the finding on the IEEE 39-bus system. The impact of  type-4 WFs, number of wind turbines, double circuit lines, FACTS devices, grid strength, evolving and cross-country faults, power swings, voltage instability, load encroachment, auto-reclosing on permanent faults, high impedance faults, CT saturation, sampling rate, window size, synchronization errors, noise, self-training, etc. are explored in Sec. VI. Comparison with other contemporary methodologies is provided in Sec. VII and the conclusions are outlined in section VIII.

\section{MODELING \& SIMULATION}

\begin{figure}[ht]\vspace{-3mm}
\captionsetup{justification=centering,textfont=normal}
\centerline{\includegraphics[width=3.53 in, height= 2.5 in]{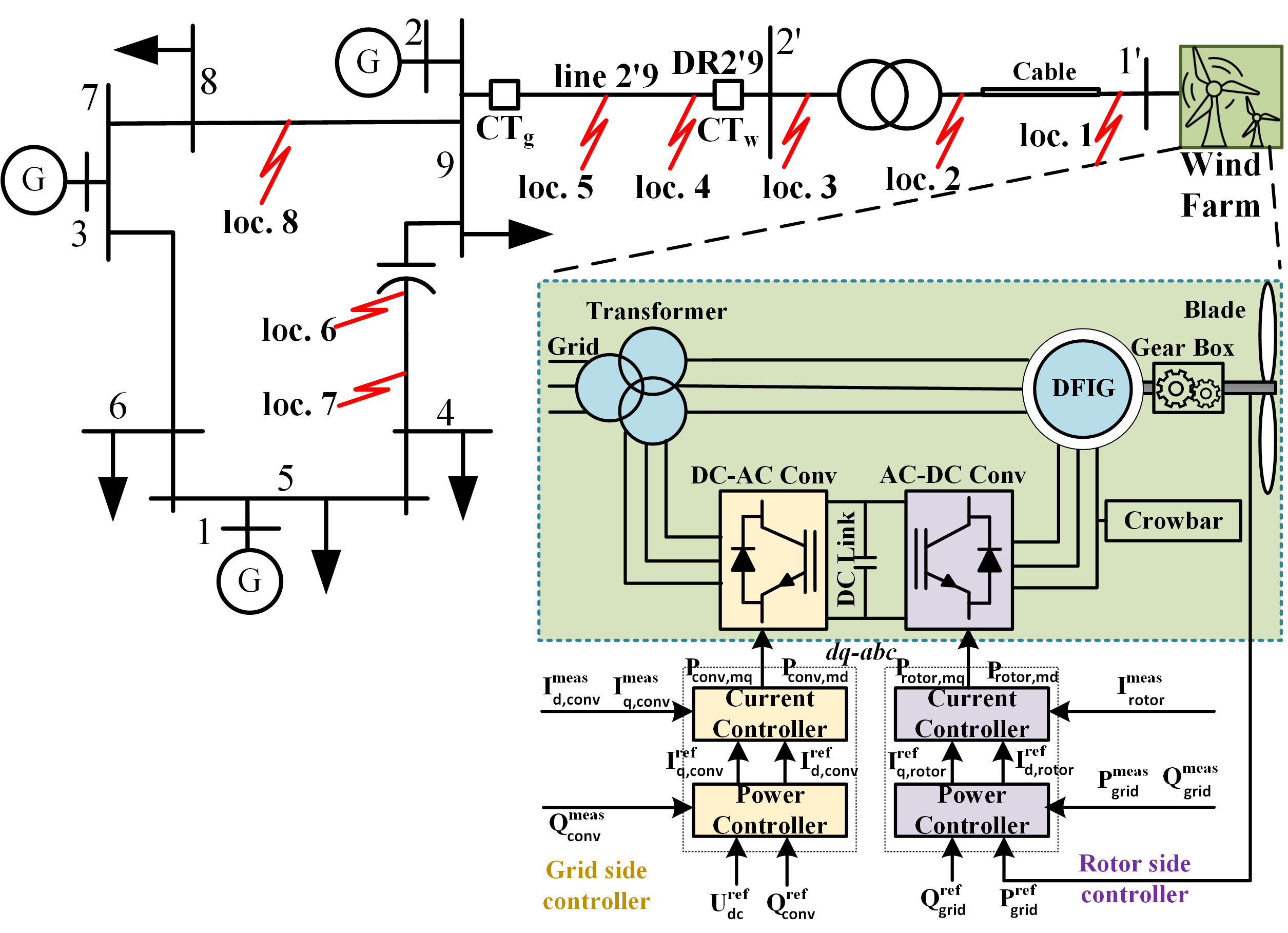}}
\caption{IEEE 9-bus test system with WF}
\label{cktsystem}
\vspace{-3mm}\end{figure}

{The IEEE 9-bus test system with a type-3 WF (Fig.\ref{cktsystem}) is used to simulate the fault and non-fault conditions.} The system modeled in PSCAD/EMTDC is described in Table \ref{system_parameters}. {A capacitor bank is connected in series with line 4-9. Each capacitor in the bank is protected by a MOV.} The faults are simulated at 8 different locations by varying fault resistance, fault inception angle, and transformer(T) connections at different generator slips(s). 
The generator at bus-2 is connected to analyze the impact of infeed in the case of capacitor and load switching at bus-5, 8, and 9. Parameters and their values used to simulate the faults, and load and capacitor switching cases are given in Table \ref{parameters1}.

\begin{table}[ht]
\centering
\renewcommand{\arraystretch}{0.95}
\setlength{\tabcolsep}{2 pt}
\caption{System Parameters}
\vspace{-1mm}
\label{system_parameters}
\footnotesize
\begin{tabular}{lll} 
\hline
\rowcolor[rgb]{0.91,0.91,0.91}  Component                                       & Parameters                     & Value                  \\ 
\hline
\multirow{5}{*}{wind farm}                                                            & capacity (single turbine)       & 2MW                    \\
                                                                                      & number of turbines                 & 100                    \\
                                                                                      & dc bus vtg. \& capacitance & 1.45kV, 7500$\mu$F      \\
                                                                                      &  AC filter                    &    700$\mu$F, 350$\mu$F, 620$\mu$H, 1.3$\Omega$                    \\
                                                                                      & rated voltage  \& frequency       & 0.9kV, 60 Hz           \\ 
\hline
\multirow{3}{*}{transmission line}                                                    & positive seq. impedance    & 0.96+31.18j$\Omega$    \\
                                                                                      & zero seq. impedance        & 33.6 + 112.9j$\Omega$  \\
                                                                                      & length  \&      voltage          & 100km, 230kV           \\ 
\hline
\multirow{2}{*}{collector cable}                                                  & positive seq. impedance    & 0.19+0.24j$\Omega$     \\
                                                                                      & zero seq. impedance        & 0.24+0.16j$\Omega$     \\  \hline
\multirow{3}{*}{\begin{tabular}[c]{@{}l@{}}transformer(T)\\(after cable)\end{tabular}} & rated capacity                 & 250MVA                 \\
                                                                                      & turn ratio                     & 33kV/230kV             \\
                                                                                      & connection                     & YNd                    \\  \hline
\multirow{3}{*}{\begin{tabular}[c]{@{}l@{}}transformer\\(before cable)\end{tabular}}     & rated capacity                 & 2.5MVA                 \\
                                                                                      & turn ratio                     & 33kV/0.69kV/0.9kV      \\
                                                                                      & connection                     & YNYNYN  \\\hline              
\end{tabular}
\vspace{-2mm}\end{table}

\begin{table}[ht]
\renewcommand{\arraystretch}{1}
\setlength{\tabcolsep}{5 pt}
\centering
\captionsetup{justification=centering}
\caption{{Simulation parameters \& values for 9-Bus} }\label{parameters1}
\vspace{-1mm}
\scriptsize
\begin{tabular}{@{}l|l@{}}
\hline
\rowcolor[rgb]{0.891,0.891,0.891} \multicolumn{2}{c}{\scshape{Fault Events} }                                              \\ \hline
Fault type             & $ag, bg, cg, ab, bc, ca, abg, bcg, cag, abcg$ (10) \\ 
Fault resistance ($R_f$)       & 0.01, 1, 10 $\Omega$  (3)                            \\
Fault inception angle  & 0$^{\circ}$, 60$^{\circ}$, 120$^{\circ}$, 180$^{\circ}$, 240$^{\circ}$, 300$^{\circ}$   (6)                       \\
Transformer connection & Y-Y, Y-$\Delta$   (2)                                 \\
Wind speed ($v$)            & 8m/s(s=0.2), 9m/s(s$\approx$0), 11m/s(s=-0.2), 22m/s(s=-0.2)   (4)                        \\
Fault location         & 1, 2, 3, 4, 5, 6, 7, 8     (8)                               \\ \hline
\rowcolor[rgb]{0.95,0.95,.95}\multicolumn{2}{l}{Total fault cases =$ 10\times 3\times 6 \times 2 \times 4 \times 8$  = 11520}                          \\ \hline
\rowcolor[rgb]{0.891,0.891,0.891} \multicolumn{2}{c}{\scshape{Non-Fault Events} }                                          \\ \hline
      Location          &     Bus-5, 8, 9    (3)                                              \\
    Switching time      &         0$^{\circ}$ to 360$^{\circ}$ in steps of 15$^{\circ}$ (25)             \\     
   Wind speed ($v$)         &         11m/s, 22m/s (2)             \\   
   Generator (at bus-2)             &         Connected/Not-connected (2)  \\   
   Capacitor/Load Rating                         &        4    \\ \hline
   \multicolumn{2}{l}{ Capacitor switching cases = $3\times 25\times 2 \times 2 \times 4$ = 1200}  \\ 
    \multicolumn{2}{l}{ Load switching cases = $3\times 25\times 2 \times 2 \times 4$ = 1200}  \\ \hline
\rowcolor[rgb]{0.95,0.95,0.95}      \multicolumn{2}{l}{Total non-fault cases = 2400}  \\ \hline
\end{tabular}
\vspace{-2mm}\end{table}

\begin{figure}[ht]\vspace{-3mm}
\captionsetup{justification=centering,textfont=normal}
\centerline{\includegraphics[width=3.53 in, height= 1.1 in]{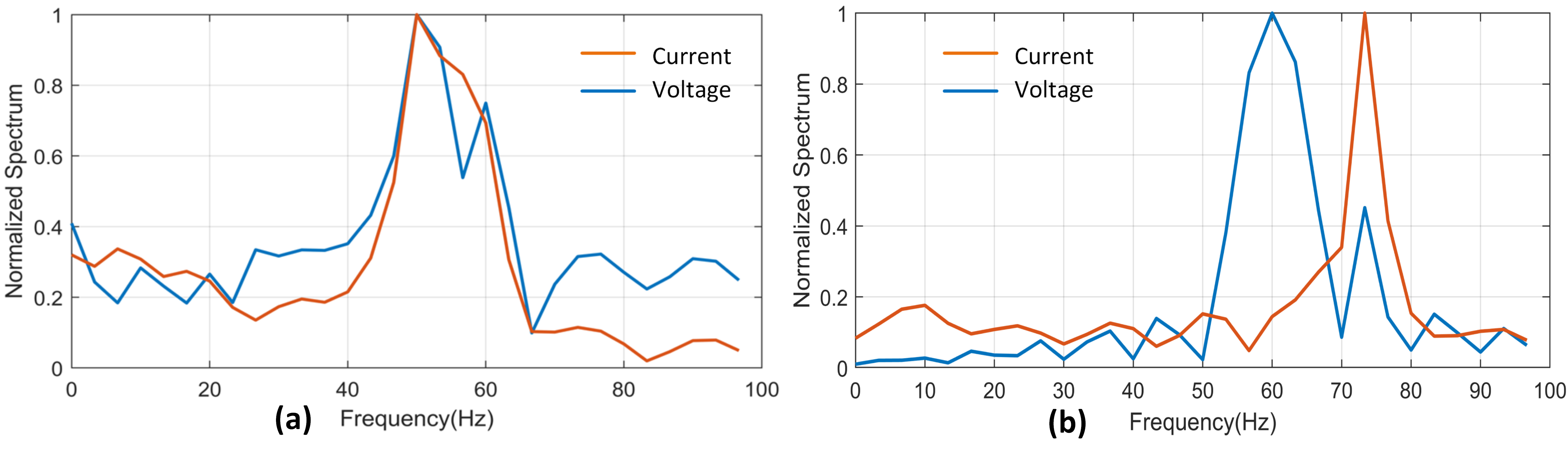}}
\caption{\small{Current and voltage spectrum for different rotor speeds (a) sub-synchronous (+ve slip) \& (b) super-synchronous (-ve slip) }}
\label{freqtrak}
\vspace{-3mm}\end{figure}

\begin{figure}[ht]\vspace{-3mm}
\vspace{-0.1cm}
\captionsetup{justification=centering,textfont=normal}
\includegraphics[width=3.53 in, height=2.3 in]{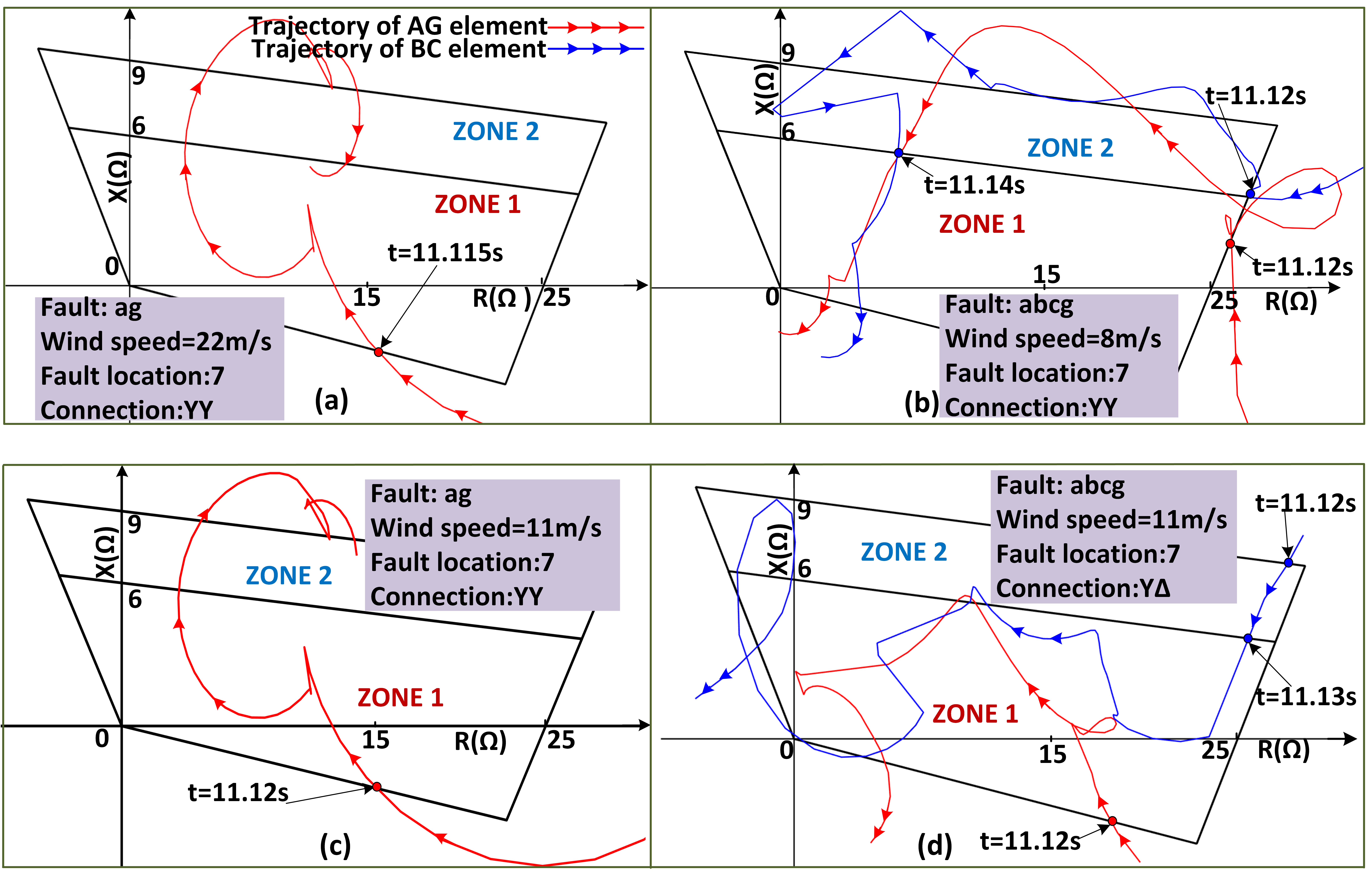}
\vspace{-0.2cm}
\caption{\small{Trajectory of AG and BC elements of the distance relay at $CT_w$ for faults at location 7 for different wind speeds, fault types, and transformer(T) connections with FIT=11.0s \& $R_f$=0.01$\Omega$}}
\label{wf3faultsmaloperate}
\vspace{-3mm}\end{figure}

\noindent \textit{Distance relay impedance trajectory:} The impedance plane of the quadrilateral distance relay DR2$'$9 connected to the WF exhibits fault current variations. A fifth-order low-pass Butterworth filter with a cut-off frequency of 480 Hz is used to avoid aliasing. Machine slip for DFIG-based WFs results in off-nominal and different frequencies for current and voltage. The frequency spectra of current and voltage phasors obtained for 8m/s (0.2 slip) and 11m/s (-0.2 slip) for 3 phase to ground ($gnd$) faults at location 5 having peaks at 50 \& 60Hz, and 72 \& 60Hz are shown in Fig.\ref{freqtrak}(a) and  Fig.\ref{freqtrak}(b) respectively. The fixed window discrete Fourier transform (DFT) measures erroneous current and voltage phasors and distance relay mal-operates. 
{Frequency tracking is used to accurately measure the phasors even during frequency excursions \cite{freqtrak}}. Currents and voltages sampled at 1920 Hz are measured using one-cycle DFT filters with a variable window size of 22-38 sample points for fundamental component extraction from individual phase voltages and currents. However, the distance relay fails to measure impedance correctly and still mal-operates.
{Impedance trajectory of} DR2$'$9 for faults at {location 7} which are outside zone 1 (80\% of line 2$'$-9) for different wind speeds, fault types, and transformer(T) connections with FIT of 11.0s and $R_f$=0.01$\Omega$ are shown in Fig.\ref{wf3faultsmaloperate}.

 \vspace{-2mm}
\section{AR Coefficient based proposed Protection }
\subsection{Proposed Protection Scheme}
\begin{figure}[ht]\vspace{-2mm}
\captionsetup{justification=centering,textfont=normal}
\centerline{\includegraphics[width=3.53 in, height= 1.9 in]{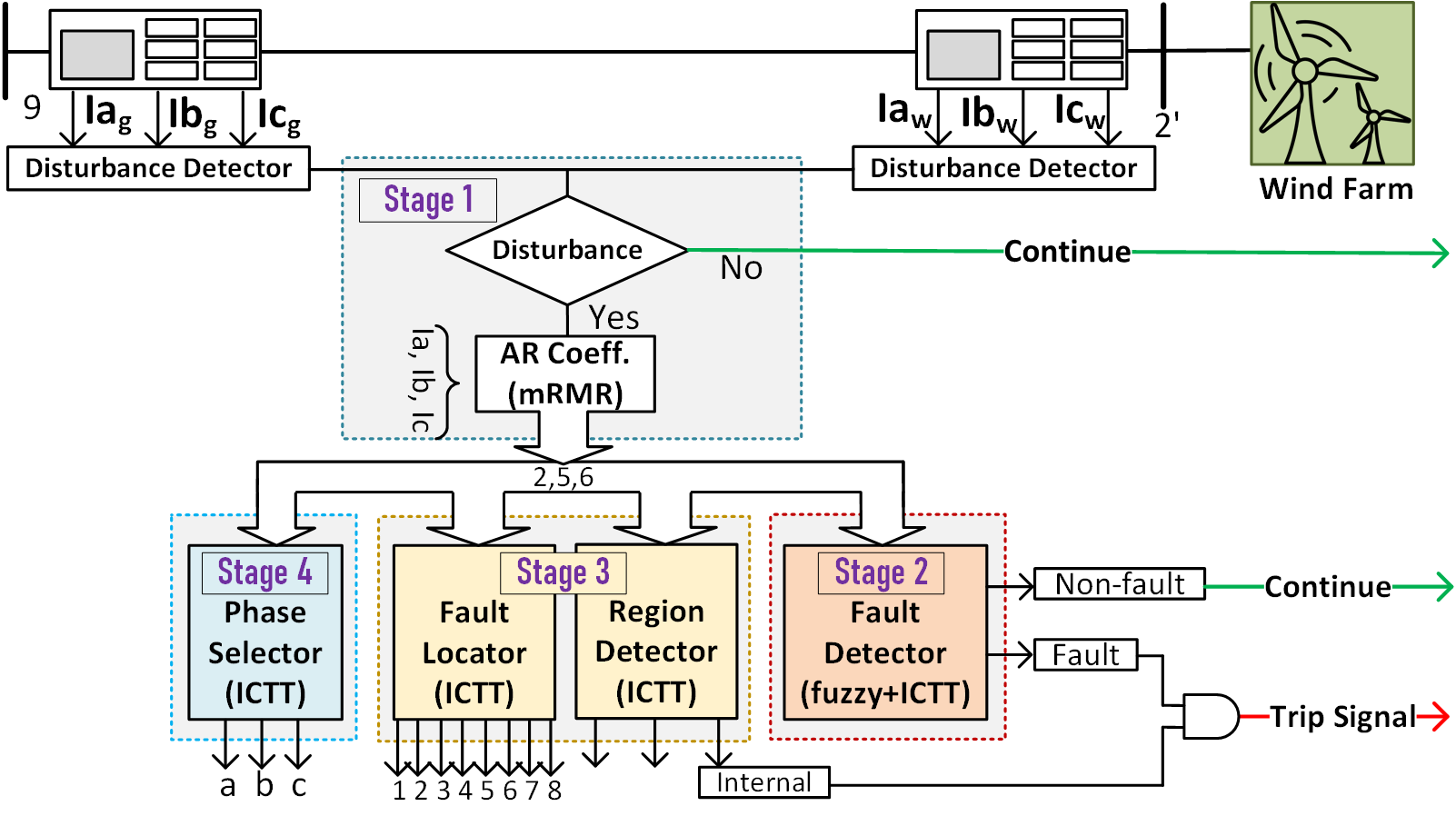}}
\caption{Flow diagram of proposed protection scheme}
\label{fc}
\vspace{-3mm}\end{figure}

{The proposed protection mechanism consists of a four-stage process illustrated in Fig.\ref{fc}}. First, the disturbance detector checks for any disturbance in the 3-phase currents measured by the CTs ($I_g$ and $I_w$) on the two sides of the TL under consideration (line 2$'$-9) and records the data in case of transients. The recorded half-cycle 3-phase currents are used to extract AR coefficients.  
The second stage incorporates the AR coefficient-based fuzzy inference system for fault detection which is supervised by the AR coefficient-based InceptionTime (ICTT) model. Third, after faults are detected, the faulty region is determined by differentiating faults among internal, forward, and backward faults. If the fault detector and fault region identifier identify the captured transient data as an internal fault, the line 2$'$-9 is instantly tripped. The exact location of the fault is also determined in this stage. Fourth, the faulty phases are identified and  different faults are classified into ten possible fault types. The first stage includes data pre-processing and feature extraction. mRMR is used to rank a list of features. In the subsequent stages, ICTT models are trained using the chosen AR coefficient feature. 
\vspace{-2mm}
\subsection{Disturbance Detector (DD)}
The disturbance detector is used to detect any change in the 3-phase currents $I_g$ and $I_w$. It computes the fractional increase between the cumulative sum of the modulus of current samples of two successive half cycles (eqn.\ref{eq_ed}).
\begin{equation}\vspace{-1mm} 
\label{eq_ed}DD (t) = \frac{\sum_{t=0}^{N_s/2}|I_{ph}(t)|-\sum_{t=0}^{N_s/2}|I_{ph}(t-N_s/2)|}{\sum_{t=0}^{N_s/2}|I_{ph}(t)|}
\vspace{-1mm} 
\end{equation}
where $N_s$ is number of samples in one cycle, and $I_{ph}$ is phase current.
The 3-phase current samples are recorded by the DD filter from the time instant  $t$ which satisfies eqn. (\ref{ed}).
\begin{equation}\vspace{-1mm}\label{ed}
{ DD(t) \geq  \beta = 0.05} \  \forall \  ph \in a,b,c  \vspace{-1mm} 
\end{equation}

The threshold $\beta$ is obtained by using the grey wolf optimizer. Grey wolf optimization (GWO) is a metaheuristic algorithm used to search for the optimal solution and is based on the social hierarchy and hunting mechanisms of a pack of wolves\cite{greywolf}. {The process is described in  algorithm \ref{gwob}.} DD(t) values  are negligibly small in the absence of transients. 

\begin{algorithm}
\small
\caption{Grey Wolf Algorithm for optimal $\beta$}\label{gwob}
{Step 1: Randomly initialize grey wolf positions \& set population size=30, search dimension=1, lower bound=0, upper bound=1,  maximum iteration=100\\
Step 2: Evaluate fitness values using the objective function:\\
 \vspace{1mm}$(1 - \frac{ No. \ of \ disturbances \ detected \ inside \ disturbance\ zone \ (1 \ cycle)}{ \ Total \ no. \ of \ disturbances\ detected})$ \\
Step 3: Iterate \& update wolf positions based on dominance hierarchy using GWO equations. Keep positions within search space. Evaluate fitness for updated positions.\\
Step 4: Track the best solution found \& its fitness value during  optimization. \\
Step 5: Return the best solution \& its fitness value after optimization.}
\end{algorithm}

\vspace{-2mm}
\subsection{Maximum Relevance Minimum Redundancy (mRMR)}
Basic information theory-based feature selection methods rank the best features exclusively on the basis of mutual information, disregarding any relationships between the chosen features. They determine which features have the highest mutual information scores between the target \textit{($y$)} and the joint distribution of all features \textit{($f_i$)}. However, the selected features might be correlated and not cover the entire feature space. Using the mutual information score and a feature's redundancy when other features are also present, mRMR penalizes a feature's relevance. It looks for features \textit{F}  that satisfy {eqn. (\ref{maxrel})} and thus selects the features with maximum mutual information \textit{I($f_i$;y)} to the target class \textit{y} and also satisfy {eqn. (\ref{minred})} in order to reduce the redundancy of the features selected using maximum relevance {(eqn. (\ref{maxrel})) }\cite {mrmr}.

\begin{equation}\vspace{-2mm} \label{maxrel}
maximum \ \mathzapf{D}(F,y), \mathzapf{D}= \frac{1}{|F|}\sum_{f_i\in F}I(f_i;y)
\vspace{-1mm}
\end{equation}
\begin{equation}\vspace{-2mm} \label{minred}
minimum \  \mathzapf{R}(F), \mathzapf{R}= \frac{1}{|F|^2}\sum_{f_i,f_j \in F}I(f_i;f_j)
\end{equation}
Above, \textit {I($f_i$;y)} \& \textit { I($f_i$;$f_j$)} are mutual information that indicate how much the joint distribution and the product of the marginal distributions of the two involved random variables differ from one another. mRMR chooses AR coefficients as the most significant from the 144 features listed in Table \ref{featuretable}.

\begin{table}[ht]
\scriptsize
\renewcommand{\arraystretch}{1.2}
\setlength{\tabcolsep}{1 pt}
\centering
\captionsetup{justification=centering}
\caption{{List of features extracted} }\label{featurelist}
\vspace{-1mm}
\label{featuretable}
\begin{tabular}{|c|c|c|} \hline
$Id$ & $Feature$ &  $Description$                \\ \hline
1           &  abs\_energy   & absolute energy,$\sum_{i=1}^{} s_{i}^2$ \\ \hline 
2           &  abs\_sum\_changes            &  absolute value of successive changes,$\sum_{i=1}^{N_s-1} \!\!|s_{i+1}\!\!-\!\!s_{i}|$ \\ \hline 
3-12           &    AR coefficient               &    $\mathbf{A}_k \forall k\!\in\! \mathbb{N}| k\leq10$    (eqn.\ref{ar})  \cite{AR}                                                                \\ \hline 
13          &   mean\_abs\_changes         &      $\frac{1}{N_s-1}\sum_{i=1}^{N_s-1} |s_{i+1}\!\!-\!\!s_{i}|$                     \\ \hline 
            
 14       & standard deviation  &  $\sigma\!\!=\!\!\sqrt{\frac{1}{N_s}\sum_{t=1}^{N_s}(s_{i}\!\!-\!\!\mu)^2}$  with mean $\mu $                                                                                  \\ \hline
  15-19          &   autocorrelation           &     $\frac{1}{(N_s-1)\sigma^2} \!\sum_{i=1}^{N_s-l}(s_i\!-\!\mu) (s_{i+l}\!-\!\mu)  \forall l\!\in\! \mathbb{N}| l\leq5 $ 
  \\ \hline   
      20              & kurtosis  &  fourth standardized moment, $\mu_4/\sigma^4$  \\  \hline

  21-100           & Fft coeff.  &  complex  $F_k\!\!=\!\!\sum_{m=1}^{N_s-1}a_m e^{(-2\pi i\frac{mk}{N_s})}\forall k\!\in\! \mathbb{N}| k\leq20$  \\ \hline
  101-130           & wavelet coeff.  & mexican hat coeff. with width of 5, 10, 20  \\ \hline
       131            &   sample entropy           &    sample entropy of time series signal  \\  \hline           
      132            & first maxima  &   first occurrence of maximum \\  \hline

       133            & last maxima  &    last occurrence of maximum    \\ \hline
       134-138            & five no. summary  &  min., 0.25 quantile, median, 0.75 quantile, and max.  \\  \hline
      139             & skewness & third standardized moment, $\mu_3/\sigma^3$   \\  \hline
     140 & variation coeff.  &   ratio of standard deviation and mean, $\mu/\sigma$ \\   \hline
        141           & complexity  & time series complexity, $\sqrt{\sum_{i=1}^{N_s-1} (s_{i+1}-s_{i})^2}$  \cite{cid}   \\  \hline
         142           & 0 seq. current  &  $I_0= {(I_a+I_b+I_c)}/{3}$ \\  \hline
           143          & + seq. current  &  $I_1= {(I_a+\alpha I_b+\alpha^2 I_c)}/{3}$, $\alpha=1\angle 120^{\circ}$ \\  \hline
            144         & -- seq. current  &  $I_2= {(I_a+\alpha^2 I_b+\alpha I_c)}/{3}$ \\  \hline

\end{tabular}
\vspace{-2mm}\end{table}

 \vspace{-2mm}

\subsection{Autoregressive (AR) coefficients}
The 3-phase relay currents can be described using relevant extracted features, and frequently, these features also reveal fresh information about the dynamics of the fault currents.
When describing time-varying phenomena, AR models which reflect random processes are often adopted in signal analysis, statistics, and finance.
AR models have served as inspiration for time series prediction in recent years. However, their application in identifying power system events is limited. AR coefficients were used to differentiate faults and non-faults in an interconnected system in \cite{systempallav}. AR coefficients were also used to distinguish faults for TL connected to WF in \cite{peci}.
Using a linear regression model, the AR process determines the present value based on the past values \cite{AR}. It uses the constant ($\mathbf{A}_0$), AR coefficients ($\mathbf{A}_{k's}$), prior observations ($s_{t-k}$), gaussian noise  ($\eta_t$) with zero mean and variance $\sigma_{\eta}^2$, and maximum lag $p$ to compute the observation $s_t$ at time $t$. 
\vspace{-2mm}
\begin{equation}\vspace{-2mm}\label{ar} s_t = \sum_{k=1}^{p} \mathbf{A}_k \ s_{t-k} + \eta_t
\end{equation}

\noindent The coefficients are calculated with the least-squares approach. Assuming that there are K samples, the coefficient  $\mathbf{A}_k$ can be estimated by arranging the K samples in a group of $L$=$K+1-p$ linear equations using (\ref{arc1}), which yields eqns.(\ref{arc2}) and (\ref{arc3}).
 \vspace{-7mm}

\begin{equation}\vspace{-3mm}
\label{arc1}
\begin{bmatrix}
s_{0} & s_{1} & \cdots & s_{p-1} \\
s_{1} & s_{2} & \cdots & s_{p} \\
s_{2} & s_{3} & \cdots & s_{p+1} \\
\vdots & \vdots & \ddots & \vdots \\
s_{K-p} & s_{K-p+1} & \cdots & s_{K-1}
\end{bmatrix} \begin{bmatrix}
\mathbf{A}_p \\
\mathbf{A}_{p-1} \\
\vdots \\
\mathbf{A}_1
\end{bmatrix} = \begin{bmatrix}
s_{p}\\
s_{p+1}\\
s_{p+2}\\
\vdots\\
s_{K}
\end{bmatrix}\vspace{-1mm}
\end{equation}

\begin{equation}\vspace{0mm}
\label{arc2}
\begin{bmatrix}
{s}_{K-p, L} & {s}_{K-p+1, L} & \cdots & {s}_{K-1, L}
\end{bmatrix} {\mathbf{A}} = \mathbf{s}_{K, L}
\vspace{-4mm}\end{equation}

\begin{equation}\vspace{-1mm}\label{arc3}{S}_{K, L} \ \mathbf{\mathbf{A}} = \mathbf{s}_{K, L}\end{equation}
where, \(\mathbf{s}_{k, L}\) represents a column vector in the matrix whose elements are the past \(L\) values of \(s_t\) which starts at the instant \(k\). 
\(\mathbf{S}_{K, L}\) consists of the columns \(\mathbf{s}_{K-p, L}, \,\mathbf{s}_{K-p+1, L}, \, \ldots \, , \mathbf{s}_{N-1, L}\).
Thus, the least-squares estimate of the AR coefficient $\mathbf{A}$,  $\hat{\mathbf{A}}$ can be provided by eqn. (\ref{ARc4}) assuming ${S}_{K, L}$ is full rank and ${S}_{K, L}^T{S}_{K, L}$ is invertible.
\begin{equation}\vspace{-2mm} \label{ARc4} \hat{\mathbf{A}} = ({S}_{K, L}^T{S}_{K, L})^{-1}{S}_{K, L}^T\mathbf{s}_{K, L}\end{equation}

\noindent 
For fault detection, localization, and identifying faulty phases, mRMR is used to select the relevant AR coefficients and lag $p$. Thus, $\mathbf{A}_2$, $\mathbf{A}_5$,  and $\mathbf{A}_6$ with p=10 are chosen as inputs.

 \vspace{-2mm}
\subsection{Deep Learning: InceptionTime (ICTT) Network}

\begin{figure}[ht]\vspace{-3mm}
\captionsetup{justification=centering,textfont=normal}
\centerline{\includegraphics[width=3.53 in, height= 2.07 in]{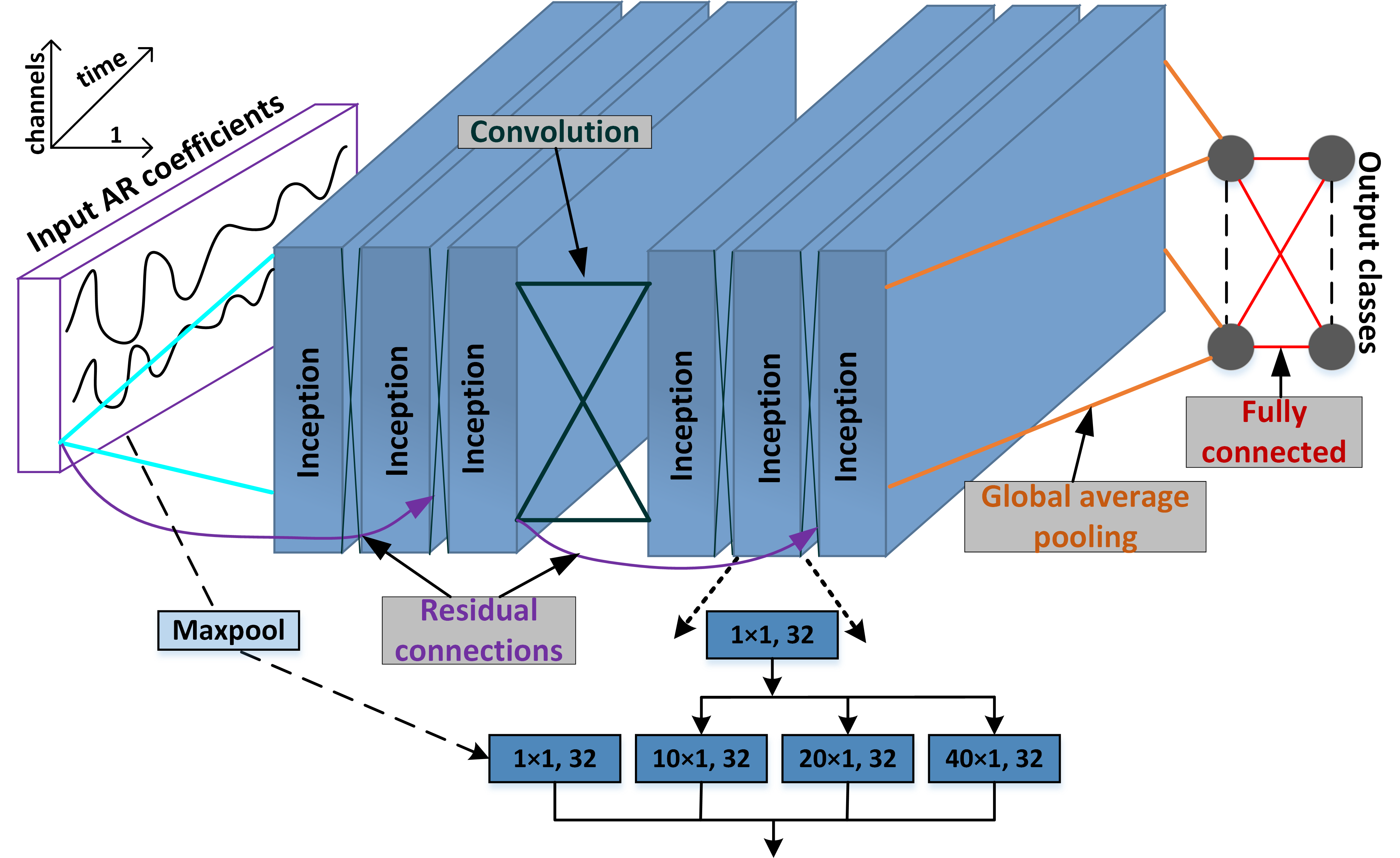}}
\caption{InceptionTime Deep Learning Network}
\label{it}
\vspace{-3mm}\end{figure}

The recent success of inception-based networks for numerous computer vision applications served as the foundation for InceptionTime (ICTT) classifier (clf.) (Fig.\ref{it}) which is a highly accurate, scalable, fast, and state-of-the-art deep learning ensemble for time series classification.
It has 2 residual blocks each having 3 inception modules instead of traditional fully convolutional layers. The basic idea of the inception module is to use different filter lengths on time series inputs parallelly extracting relevant patterns.
The gradient vanishing problem is alleviated since each residual block's input is passed to the next block's input via a linear link.  The residual blocks are followed by a global average pooling layer which averages the multivariate time series output. With softmax activation function, a final layer that is fully connected and having number of neurons indicating the number of classes is used. 

In this study, the ICTT network described in \cite{Ismail_Fawaz_2020} which consists of 5 unique randomly initialized inception networks, each created by cascading numerous inception modules and using the concept of receptive field for temporal data, is used. Eqn. (\ref{iteq}) makes use of the capabilities of these 5 ensembled inception networks rather than just one improving the standard deviation in performance. 
\begin{equation}\vspace{-2mm}
\label{iteq}  
y_{p,c}=\frac{1}{n}\sum_q^n\sigma_c(x_p, \theta_q) \ \forall  c \in \mathbb{N}| c\leq C
\end{equation}
where the logistic output $\sigma_c$ averaged across n random models signifies the output probability that the transient (fault or non-fault event) $x_p$ belongs to class c with C=2 for fault detection, C=3 for region identification, C=8 for fault location, C=7 for phase selection, and C=10 for fault classification.
The inception network is depicted in Fig. \ref{it} with six inception modules stacked one on top of the other. It also shows details of a single inception module. A bottleneck layer reduces the dimension of the input AR coefficients ($\mathbf{A}_2$, $\mathbf{A}_5$,  and $\mathbf{A}_6$) by the operation of sliding filters. In addition, a maxpooling operation followed by a bottleneck layer is used to reduce the dimension parallelly. Each layer uses 3 sets of filters having 32 filters each with lengths 10, 20, and 40 and 32 maxpooling filters. Lastly, the output of the multivariate time series is produced by joining the results of each independent parallel convolution or maxpooling. 
For each inception module in the network, the same procedures are repeated. Adam optimization is used to train the models, and Glorot uniform initializer is used to generate random numbers for all weights \cite{Ismail_Fawaz_2020}.

\section{Results on IEEE 9-Bus System}
The fuzzy-based protection system is described and findings of fault detection, fault localization, and phase selection stages of the 4-stage process on the IEEE 9-bus system are discussed in this Section.

\subsection{Evaluation Metrics for Classification Algorithms}
Analyzing the consistency with which a data point is correctly classified is one way to evaluate the performance of a classification algorithm. Accuracy is the proportion of data points that are correctly anticipated out of all the data points given by eqn. (\ref{acc}) where T is true, F is false, P is positive, and N is negative.
 \begin{equation}\vspace{-1mm}\footnotesize
\label{acc}{\eta}=\frac{TP+TN}{TP+FN+TN+FP}\end{equation}

\noindent Confidence interval (CI) of the predictive accuracy of models is also obtained to give more information with the assumption that predictions follow normal distribution \cite{confidence}.
The lower and upper limits at a confidence are thus computed using eqn. (\ref{ci}). 
\begin{equation}\vspace{-2mm}
\label{ci}
   CI= \eta \pm  Z\sqrt{\frac{1}{N_t}{\eta (\eta -1)}}
\end{equation}
Here, $Z\!\! =\!\! 1\!\! − \!\!\alpha/2$ is quantile of a standard normal distribution with $\alpha$ being the error quantile, $N_t$ denotes the total number of test cases, and $\eta$ denotes the classification accuracy. For a typical CI of 95\%, $\alpha$=0.05 and $Z$=1.96. 

\subsection{Fault detection, localization, \& classification}
\subsubsection{Fault Detection}
After data pre-processing, the second and most important stage of this all-encompassing protection strategy is fault detection. The fault detection model is primarily fuzzy-based adaptive protection while the ICTT network augments and backs up its performance.

\begin{figure}[H]
\vspace{-0.3cm}
\centering
\captionsetup{textfont=normal}
\includegraphics[width=3.53 in, height=1.4 in]{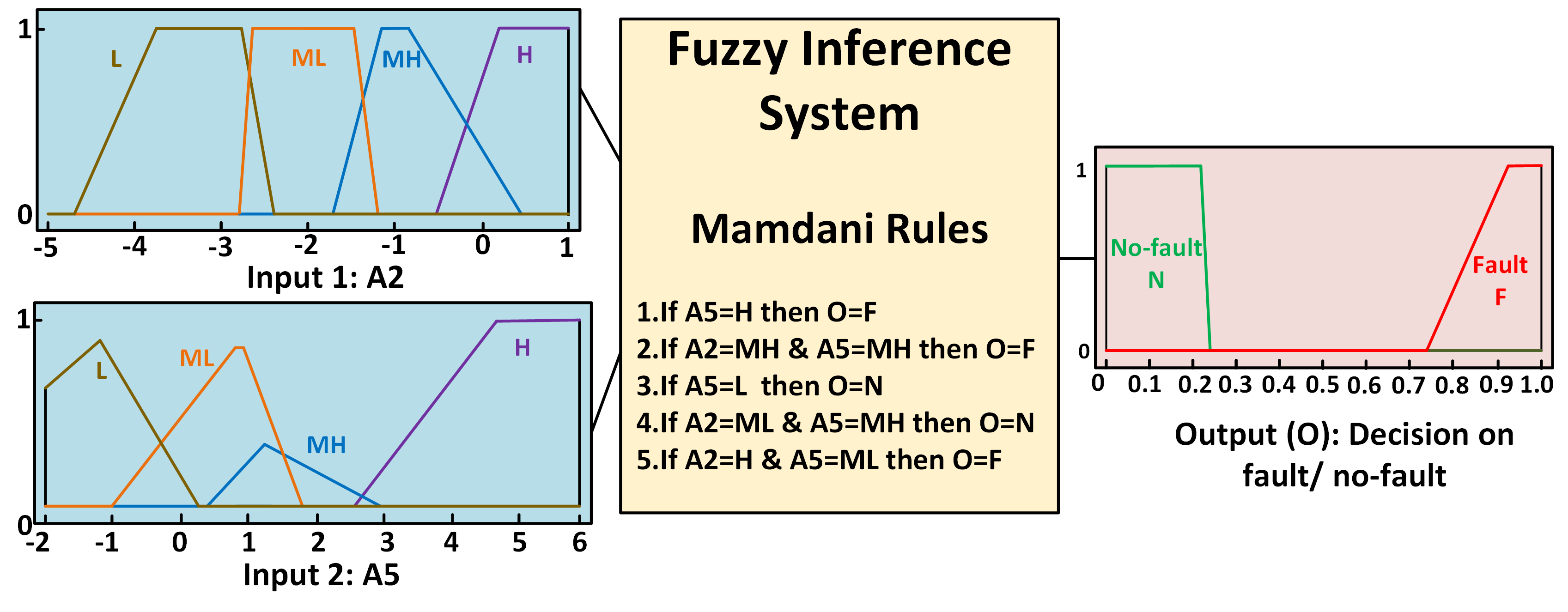}
\caption{Existing trapezoidal input \& output membership functions and rules of the fuzzy inference system for fault detection}
\label{FUZZY1}
\end{figure}

\paragraph{Fuzzy logic-based adaptive protection}
Power system operating conditions vary and an algorithm based on a fixed threshold may not be suitable for all scenarios. Fuzzy logic has been used in applications where it needs to adapt to changing system parameters. In this study, a fuzzy inference system tuned by a data-driven approach is employed to detect faults.
The parameters of input and output membership functions and the rules are intelligently tuned. Instead of trying every combination, the optimal solution is obtained by sampling a small subset of entire solution space using Genetic Algorithm. 
For fuzzy logic, maximum $\mathbf{A}_2$ \& $\mathbf{A}_5$ coefficients from  half-cycle  a, b, c phase currents at the WF end are defined as inputs. 
The adaptive protection algorithm is presented below, and the fuzzy-based decision-making system is shown in Fig. \ref{FUZZY1}.

\begin{algorithm}
\small
\caption{{Fuzzy-based adaptive protection}}\label{adaptivefuzzy}
Step 1: Extract local minimum and local maximum values of $\mathbf{A}=[\mathbf{A}_2, \mathbf{A}_5]$ calculated on half-cycles where DD detects a transient

\hspace{5mm}$\mathbf{A}_{min} = min (\mathbf{A}1, \mathbf{A}2, . . . . . . , \mathbf{A}  {N_h)} $

\hspace{5mm}$\mathbf{A}_{max} = max (\mathbf{A}1, \mathbf{A}2, . . . . . . , \mathbf{A} N_h) $ \\
where $N_h$=no. of transient half cycles in a defined time frame

Step 2: If $\mathbf{A}_{min} < \mathbf{A^-}_{min}$ or $\mathbf{A}_{max} > \mathbf{A^-}_{max}$ compute new fuzzy rules and parameters where, $\mathbf{A^-}$ = $\mathbf{A}$ for previous time frame.
\end{algorithm}

\begin{minipage}{1.74in}
\footnotesize
\raggedleft

\renewcommand{\arraystretch}{1.0}
\setlength{\tabcolsep}{2 pt}
\captionof{table}{Results of supervisory Fault Detection}
\vspace{-1mm}
\label{fd}
\footnotesize

\begin{tabular}{lccc}
\hline
\rowcolor[rgb]{0.91,0.91,0.91} clf. & $\eta$(\%)   & CI & $\eta$ SMOTE   \\
\hline
ICTT  & \textbf{99.6}  &   99.9 -99.3 & 99.7       \\
LSTM            &  96.7  &  97.5 -95.9 &  98.8       \\
FCN             & 99.2    & 99.6 -98.8 & 99.3    \\
ResNet          &  99.2   &  99.6 -98.8 & 99.5    \\
MLP              &  97.0    & 97.7 -96.3 & 97.6  \\
GRU              &  97.0   & 97.7 -96.3 & 98.7     \\ 
TST              & 97.3 &  98.0 -96.6 & 99.0   \\ \hline
\end{tabular}
\vspace{3mm}
\end{minipage}
\begin{minipage}{1.51in}
\renewcommand{\arraystretch}{1.0}
\setlength{\tabcolsep}{3 pt}
\centering
\footnotesize
\captionof{table}{Results of Region Identification}\label{fr}
\vspace{-1mm}
\begin{tabular}{lccc}
\hline
\rowcolor[rgb]{0.91,0.91,0.91} clf. & $\eta$(\%)  & CI \\ \hline
ICTT  & \textbf{99.0}   & 99.5 -98.5 \\
LSTM            & 96.5    & 97.4 -95.6 \\
FCN             & 98.6    & 99.2 -98.1\\
ResNet          & \textbf{99.0}    & 99.5 -98.5\\
MLP              &      96.4 & 97.3 -95.5 \\
GRU              &      97.2  & 98.0 -96.4 \\ 
TST              &      98.1  & 98.7 -97.5 \\ \hline
\end{tabular} 
\vspace{3mm}
\end{minipage}

\begin{figure}[ht]\vspace{0mm}
\captionsetup{justification=centering,textfont=normal}
\centering
\centerline{\includegraphics[width=3.5 in, height= 1.15 in]{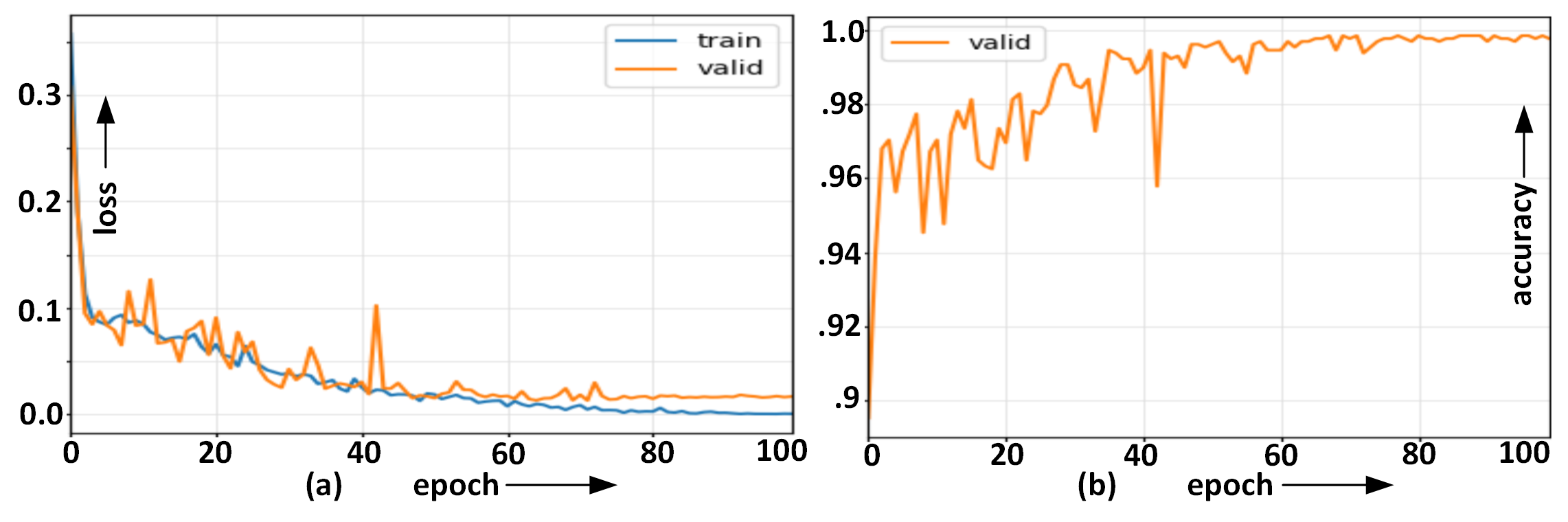}}
\caption{Learning curve:(a)loss vs epoch, (b)accuracy vs epoch}
\label{itime}
\vspace{-0.1cm}
\end{figure}

\paragraph{Deep-learning based supervisory protection}
In research articles on power system protection, many ML algorithms have shown promising outcomes.  In this study, different traditional and non-traditional ML methods are tested. Performance of deep learning models such as long short-term memory (LSTM)\cite{LSTM}, fully convolutional network (FCN) \cite{ResNetFCNMLP}, gated recurrent units (GRU) \cite{GRU}, time series transformer (TST)\cite{tstmvp}, residual network (ResNet) \cite{ResNetFCNMLP}, multi-layer perceptron (MLP) \cite{ResNetFCNMLP}, and InceptionTime (ICTT); and non-deep learning models such as 
decision tree (DT), random forest (RF), extreme gradient boosting (XGBoost),
naive bayes (NB), support-vector machines (SVM), and k-nearest neighbors (kNN) are evaluated for supervising the fault detection.
It is observed that ICTT outperforms the others with an $\eta$ of 99.6\%, leaving behind ResNet with 99.2\% and XGBoost with 99.3\%.
These classifiers take as inputs the three AR coefficients — 2, 5, and 6 — that were chosen using the mRMR algorithm from a, b, and c phase currents. 
The number of fault events (11520) and non-fault events (2400) results in an imbalanced dataset, with faults and non-faults forming the majority and minority classes respectively.  
In such cases, results may be skewed if there is not enough information available on the minority class. Synthetic minority over-sampling technique (SMOTE) is employed to counteract any potential effects that an unbalanced dataset might have \cite{smote}. Results of the supervisory fault detection with deep learning models are shown in Table \ref{fd} with and without SMOTE. The learning curve (loss curve and $\eta$) for the ICTT supervisory fault detection model is illustrated in Fig.\ref{itime}, where the x-axis denotes the number of epochs and the y-axis, loss and  $\eta$, respectively. The training and validation loss lowering to a point of stability and having a small gap between the final values demonstrates the model's good fit.

\textit{Relevance of AR-coefficients:}
A non-WF dataset available on IEEE data-port and contributed in \cite{accesspallav} is analyzed to determine the significance of the AR-coefficients for the problem of fault detection in TLs connected to WFs at hand. The authors reported an $\eta$ of 99.8\% on a two-class fault detection problem with 46872 faults and 13680 non-fault cases.
The same dataset was trained and tested with AR coefficients as inputs, however, $\eta$ of just 94.8\% establishes that the AR coefficients perform extremely well ($\eta$ of 99.6\%) for fault detection in TLs connected to WFs, but it may not be true for systems unrelated to WFs.

\begin{minipage}{1.6in}
\vspace{2mm}
\footnotesize
\captionof{table}{Performance for Fault Location}\label{fl}
\vspace{-2mm}
\begin{tabular}{lccc}
\hline
\rowcolor[rgb]{0.91,0.91,0.91} clf. & $\eta$(\%)  & CI \\ \hline
ICTT  & \textbf{96.2} & 97.10 - 95.30\\
LSTM            & 89.8  & 91.23 - 88.37   \\
FCN             & 95.1  & 96.12 - 94.08  \\
ResNet          & 95.9  & 96.83 - 94.97  \\
MLP              &      87.6 & 89.15 - 86.05 \\
GRU              &      90.3 & 91.70 - 88.9 \\ 
TST              &      94.3 & 95.39 - 93.21 \\  \hline
\end{tabular}
\vspace{3mm}
\end{minipage}
\hspace{2mm}
\begin{minipage}{1.6in}
\vspace{2mm}
\footnotesize
\captionof{table}{Performance for Phase Detection}\label{fs}
\vspace{-2mm}
\footnotesize
\begin{tabular}{lccc}
\hline
\rowcolor[rgb]{0.91,0.91,0.91} clf. & $\eta$(\%)  & CI \\ \hline
ICTT  & \textbf{98.5} & 99.1 - 97.9\\
LSTM            & 91.7  &   93.0 - 90.4\\
FCN             & 96.6   & 97.5 - 95.7\\
ResNet          & 97.1    & 97.9 - 96.3\\
MLP              &      95.6 &  96.6 - 94.6\\
GRU              &      91.1  &  92.4 - 89.8\\ 
TST              &      96.5  & 97.4 - 95.6\\ \hline
\end{tabular} 
\vspace{3mm}
\end{minipage}
\subsubsection{Fault localization}
{After identifying faults, the scheme determines the fault region and location. These faults are simulated at 8 different locations shown in Fig.\ref{cktsystem} where faults at loc.4, and loc.5 are internal; loc.1, loc.2, and loc.3 are external faults in the reverse direction; and loc.6, loc.7, and loc.8  are external faults in the forward direction of relay DR2$'$9}. The reverse and forward external faults are grouped on basis of whether the grid or the WF feeds the fault. Table \ref{fr} displays the outcomes of the deep learning models used to differentiate the internal and external faults. The line 2$'$-9 is tripped for internal faults.
Further, the fault locations are identified for the 8 {different} locations, with the outcomes summarized in Table \ref{fl}. These results show excellent performance for fault location.

\subsubsection{Phase Selection}
\begin{figure}[ht]\vspace{-3mm}
 \captionsetup{justification=centering,textfont=normal}
\centerline{\includegraphics[width=3.54 in, height= 1.7 in]{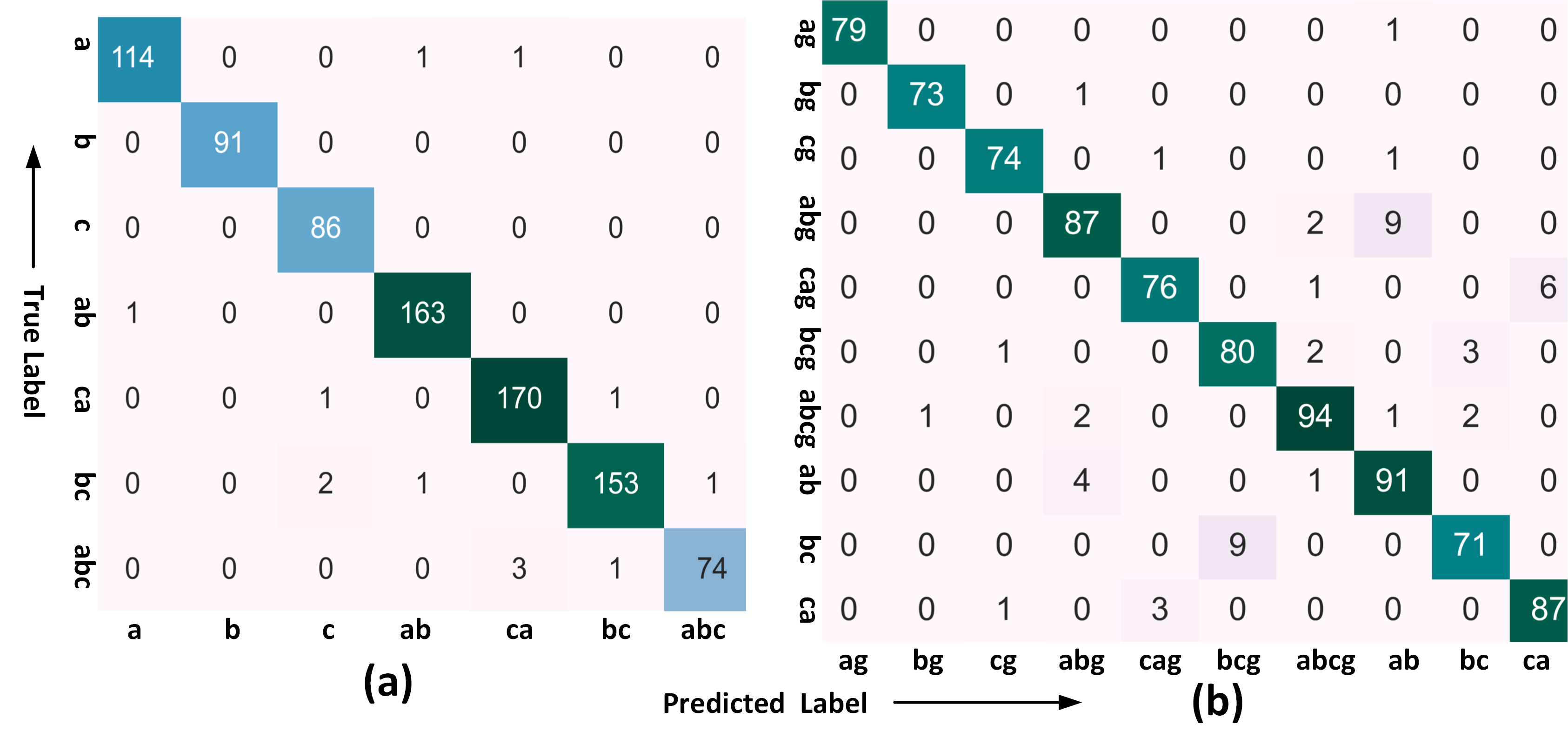}}
\caption{Confusion matrix for (a) phase selection, and (b) fault classification with ICTT}
\label{conmatrix}
\vspace{-3mm}\end{figure}

After fault detection and localization, the faulty phase is identified. The faulty phases among a, b, c, ab, bc, ca, and abc are determined.
The results of the phase selection with deep learning models are shown in Table \ref{fs}. In addition, the faults are classified into ten different fault types. The ICTT model gives an $\eta$ of 93.98\% for fault classification. Fig.\ref{conmatrix} shows the confusion matrix for the phase selection and fault classification for ICTT.

Results of fault detection, fault location, and phase selection stages with traditional classifiers are shown in Table \ref{traditional}.

\begin{table}[ht]
\centering
\footnotesize
\renewcommand{\arraystretch}{1.0}
\setlength{\tabcolsep}{4 pt}
\caption{Performance with Traditional Classifiers}
\vspace{-1mm}
\label{traditional}
\arrayrulecolor{black}
\begin{tabular}{llcllll}
\hline
\rowcolor[rgb]{0.91,0.91,0.91} \multicolumn{1}{c}{{\cellcolor[rgb]{0.91,0.91,0.91}}}                                 & \multicolumn{6}{c}{\textit{Classifier}}                       \\ 
\hhline{>{\arrayrulecolor[rgb]{0.91,0.91,0.91}}->{\arrayrulecolor{black}}------}
\rowcolor[rgb]{0.91,0.91,0.91} \multicolumn{1}{c}{\multirow{-2}{*}{{\cellcolor[rgb]{0.91,0.91,0.91}}\textit{Model}}} & \multicolumn{1}{c}{RF} & XGBoost & DT   & KNN  & NB   & SVM   \\ 
\hline
\textit{Fault Detection}                                                                                                   & 96.2                   & 99.3    & 87.0 & 99.2 & 78.4 & 94.2  \\
\textit{Region Identification}                                                                                            & 95.8                   & 98.3    & 85.4 & 94.0 & 57.3 & 92.5  \\
\textit{Fault Location}                                                                                                    & 94.8                   & 95.4    & 84.1 & 93.4 & 56.4 & 91.5  \\
\textit{Phase Selection}                                                                                                   & 95.3                   & 96.2    & 85.2 & 95.6 & 53.0 & 94.0  \\
\hline
\end{tabular}
\vspace{-2mm}\end{table}

\vspace{-2mm}
\section{Validation on IEEE 39-Bus Test System}

\begin{figure}[ht]\vspace{-2mm}
\vspace{-0.1cm}
\centering
\captionsetup{justification=centering,textfont=normal}
\includegraphics[width=3.3 in, height=3.2 in]{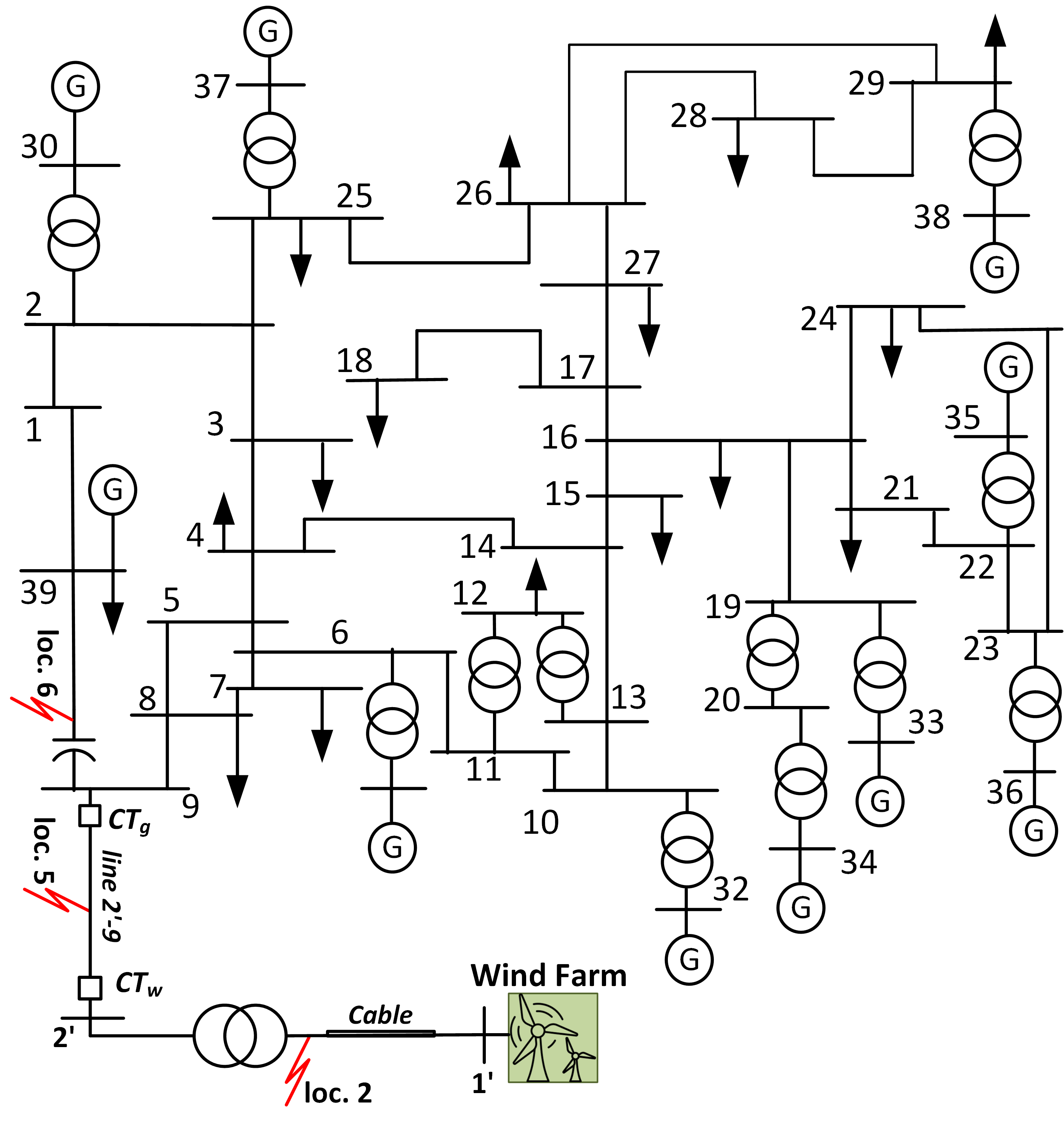}
\vspace{-0.1cm}
\caption{IEEE 39-bus with Type-3 WF at bus-9}
\label{39busSYSTEM}
\vspace{-0.1cm}
\end{figure}

The validity of the suggested scheme is evaluated on the IEEE 39-bus system (Fig.\ref{39busSYSTEM}).
Table \ref{parameters2} shows the parameters and their values used to simulate different fault cases. The AR coefficient-based supervisory fault detection applies to the 39-bus system with an $\eta$ of 99.5\% on 2160 faults and 2400 non-faults cases. The ICTT also identifies the region of fault (internal/external forward/external reverse) with 98.5\% $\eta$. The line 2$'$-9 is tripped for internal faults.

\begin{table}[ht]
\centering
\renewcommand{\arraystretch}{0.95}
\setlength{\tabcolsep}{2 pt}
\captionsetup{justification=centering}
\caption{{Fault simulation parameters for 39-Bus} }\label{parameters2}
\vspace{-1mm}
\footnotesize
\begin{tabular}{@{}ll@{}}
\hline
\rowcolor[rgb]{0.91,0.91,0.91} \multicolumn{2}{c}{\textsc{Fault Events}}                                               \\ \hline
Fault type             & ag, bg, cg, ab, bc, ca, abg, bcg, cag, abcg (10) \\ 
Fault resistance       & 0.01, 1, 10 $\Omega$  (3)                            \\
Fault inception angle  & 0$^{\circ}$ , 60$^{\circ}$ , 120$^{\circ}$ , 180$^{\circ}$ , 240$^{\circ}$ , 300$^{\circ}$    (6)                       \\
Transformer connection & Y-Y, Y-$\Delta$   (2)                                 \\
Wind speed  ($v$)           &  11m/s, 22m/s   (2)                        \\
Fault location         & 2, 5, 6   (3)                               \\ \hline
\rowcolor[rgb]{0.91,0.91,0.91}\multicolumn{2}{l}{Total fault cases =$ 10\times 3\times 6 \times 2 \times 2 \times 3$  = 2160}                          \\ \hline
\end{tabular}
\vspace{-3mm}\end{table}

\section{Discussions}
 Scenarios that could pose a challenge to the functioning of the proposed AR coefficient-based scheme are explored in the next paragraphs. {Impedance trajectories} of the conventional distance relays in such scenarios is also illustrated at the end of this section. For the sake of brevity, just the second stage of the four-stage process—supervisory fault detection—will be considered for validation.  

\subsection{Impact of a Type-4 Wind Farm}
In place of the DFIG-based WF, a permanent magnet synchronous machine with a full-scale converter is connected to the IEEE 9-bus system. The stability of the proposed scheme with type-4 WTG is validated by no misclassifications on 2160 fault and 2400 transient disturbance cases. The faults are simulated by varying $v$ (7 and 23m/s), $R_f$ (0.01, 1, and 10$\Omega$), 10 fault types, YY \& Y$\Delta$ transformer connections, and 6 fault inception times (FIT) at locations 2, 5, and 6. This study can also be used for systems that use photovoltaic generation as the grid side interfaces are similar \cite{pv}.

{\subsection{Joint impact of Type-3 \& Type-4 Wind Farms}
The type-3 WF is connected at bus-9 and the type-4 WF is connected at bus-4 in the IEEE 9-bus system. The stability of the proposed scheme with type-3 and type-4 WTG is validated by an accuracy of 99.7\% on 2160 fault and 2400 transient disturbance cases. The faults are simulated by varying $v$ (8, 9, 11,  and 22m/s), $R_f$ (0.01, 1, and 10$\Omega$), 10 fault types, and 6 FIT at locations 2, 5, and 6.}

{\subsection{Joint impact of two Type-3 Wind Farms}
One type-3 WF is connected at bus-9 and another type-3 WF is connected at bus-4 in the IEEE 9-bus system. The stability of the proposed scheme with two type-3 WTGs at different locations is validated by an accuracy of 99.8\% on 2160 fault and 2400 transient disturbance cases. The faults are simulated by varying $v$ (8, 9, 11, and 22m/s), $R_f$ (0.01, 1, and 10$\Omega$), 10 fault types, and 6 FIT at locations 2, 5, and 6.}

\subsection{Performance Evaluation for change in Wind Farm Units}
The capacity of the WF is changed by setting the wind turbine units to 70 and 130 instead of 100 and the proposed scheme is verified. The scheme identified the faults from transients with 98.7\% $\eta$ on 2400 transients and 3960 faults obtained by varying $v$ (8 and 11m/s), $R_f$ (0.01, 1, and 10$\Omega$), 10 fault types, and 11 FIT at locations 2, 5, and 6. 
\subsection{Performance in case of Weak Grid}
When WFs are integrated with weak systems, differential protection could fail in the event of line faults due to a larger restraining current \cite{pearson18}. To test the performance of the proposed scheme under different grid strengths the short circuit ratio (SCR) is changed from 12.9 (strong grid) to 2.4 (weak grid) by reducing the number of generator units (120 MVA each) at bus-9 from 8 to 1. SCR in this case is the ratio of the short circuit MVA at bus-9 to the MW rating of the WF. 
The scheme identified the faults from transients with 99.0\% $\eta$ on 2400 transients and 1080 faults obtained by varying $v$ (9 and 11m/s), $R_f$ (0.01, 1, and 10$\Omega$), 10 fault types, and 6 FIT at locations 2, 5, and 6.

\subsection{Impact of Double Circuit Line}
Mutual coupling of double circuit TLs poses a threat to the reliability of ground distance relays, necessitating special consideration \cite{double}. A double-circuit TL of 230kV, 100km, and 60Hz is connected between bus-2$'$ and 9. High current flows in the protected line when a fault occurs in the parallel line, confusing the protective system. 
The proposed scheme identified the faults from transients with 99.7\% $\eta$ on 2400 transients and 2520 faults simulated by varying $v$ (8, 9, 11, and 22m/s), $R_f$ (0.01, 1, and 10$\Omega$), 10 fault types, and 21  (FIT) at the middle of this line.

\subsection{Impact of FACTS devices}
\subsubsection{TCSC} TCSCs used with WFs to enhance power transfer may affect the performance of conventional distance relays \cite{sauviktcsc21}.  A TCSC having a capacitor, an inductor, and MOV for overvoltage protection providing maximum compensation of 50\% is placed between bus-9 and 4 in the 9-bus test system (Fig.\ref{tcscfig}a). 2400 non-faults and 2160 faults are simulated by varying $v$ (8, 9, 11, and 22m/s), location (2, 5, and 7), resistance (0.01, 1, and 10$\Omega$), 10 fault types, 6 FIT with the TCSC placed at location 7. The proposed scheme identified the faults with 97.2\% $\eta$.

\subsubsection{Phase Shifting Transformers (PST)} 
\begin{figure}[ht]\vspace{-1mm}
\vspace{-0.1cm}
\centering
\captionsetup{justification=centering,textfont=normal}
\includegraphics[width=3.5 in, height=1.4in]{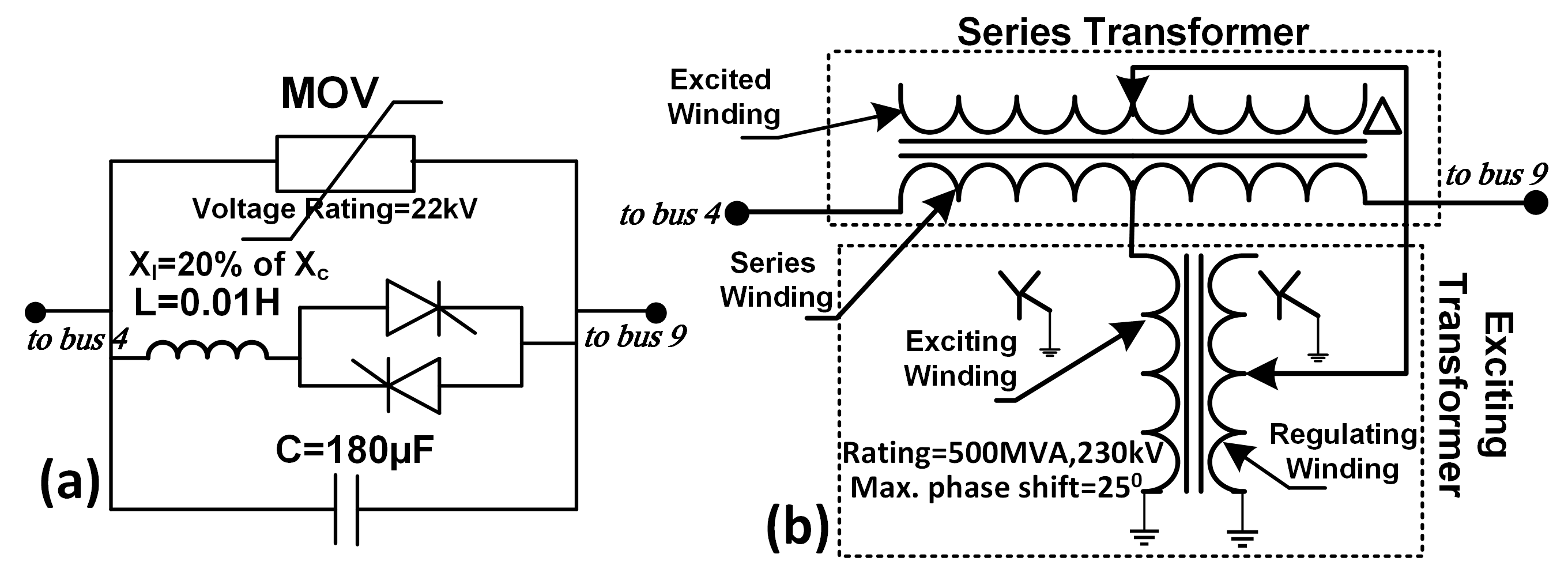}
\vspace{-0.2cm}
\caption{Configuration:(a)TCSC, (b)Indirect Symmetrical PST}
\label{tcscfig}
\vspace{-0.1cm}
\end{figure}

PSTs offer active power flow control, enhance grid flexibility and stability, and assist operators in making the most of the current infrastructure \cite{accesspallav}. The positioning of CTs and VTs is crucial for the protection of lines compensated with phase shifters. 
A 500MVA, 230kV rated indirect symmetrical PST offering a maximum phase shift of $\pm25^{\circ}$ connects buses 9 and 4 (Fig.\ref{tcscfig}b).
With 2160 PST faults obtained at 0.3 tap position by varying phase shift (backward \& forward), $v$ (8 and 11m/s), $R_f$ (0.01, 1, and 10$\Omega$), 10 fault types, 6 FIT at fault locations 2, 5, and 6 and 2400 non-fault transients, the proposed scheme identified the faults from other transients with no errors.

\subsection{Impact of Cross-country \& Evolving Faults}

 \noindent Cross-country faults, which involve simultaneous faults at two different locations (Fig.\ref{faults}a,b), and evolving faults, in which a primary fault and a secondary fault occur at the same location but at different FIT (Fig.\ref{faults}c), both have a negative impact on the effectiveness of the distance relaying scheme \cite{crossevolve}.

The proposed scheme is evaluated for 1188 cases of cross-country faults obtained by varying 11 FIT, $R_f$ (0.01, 1, and 10$\Omega$), and $v$ (8 and 11m/s). Simulated cases include simultaneous $lg$ faults at location 7 and location 5 (e.g. $bg$  at loc. 7 and $ag$, $bg$, $cg$ at loc. 5) (Fig.\ref{faults}a) and simultaneous $lg$ faults in ckt.1 and ckt.2 (e.g. $cg$ in ckt.1 and $ag$, $bg$, $cg$ in ckt.2) at same location 5 (Fig.\ref{faults}b) {of the double circuit line}. The protection scheme identified the faults with no errors.

The proposed scheme is also evaluated for 396 cases of evolving faults obtained by varying 11 FIT, $R_f$ (0.01, 1, and 10$\Omega$), and $v$ (8 and 11m/s). $lg$ faults are converted to $llg$ faults (e.g. $ag$ to $abg$, $acg$; $bg$ to $abg$, $bcg$; $cg$ to $bcg$, $acg$) at location 5 (Fig.\ref{faults}c). The scheme identified the faults with no errors.

\subsection{Discrimination of Close-in and Remote-end faults}
\begin{figure}[ht]\vspace{-1mm}
\vspace{-0.1cm}
\centering
\captionsetup{textfont=normal}
\includegraphics[width=3.5 in, height=2.1 in]{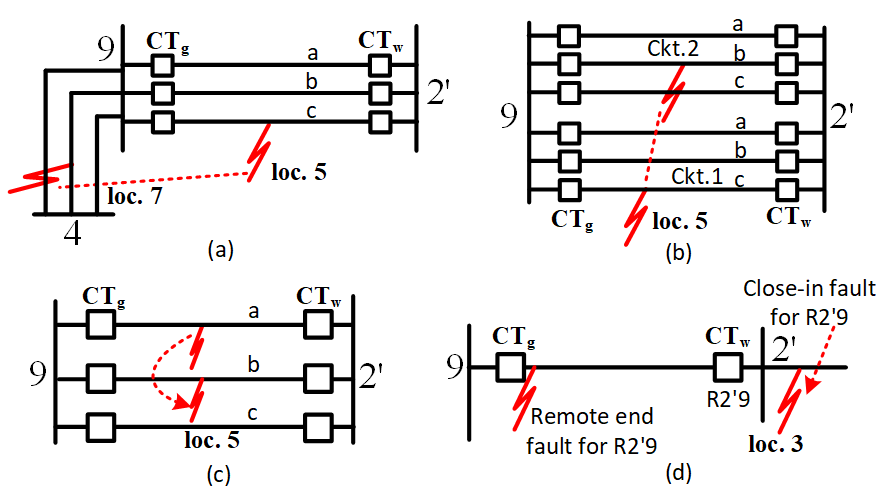}
\vspace{-0.3cm}
\caption{{(a) Cross-country: bg at loc.7 \& cg at loc.5 at 11.0s, (b) Cross-country: cg (ckt.1) \& bg (ckt.2) at loc.5 at 11.0s, (c) Evolving: ag at 11.0s converted to abg at 11.008s at loc.5, and (d) Close-in \& remote-end faults}}
\label{faults}
\vspace{-0.1cm}
\end{figure}

Conventional relays may malfunction for close-in faults due to low voltage and high current magnitude causing the CT to saturate, and remote faults may be challenging to detect because the voltage and current magnitude are within normal operating range. 
720 faults simulated at locations 5 and 3 are used as remote-end and close-in faults (Fig.\ref{faults}d) by varying FIT, $R_f$, $v$, and fault types. With AR coefficients of 3-phase currents at ${CT_w}$ as inputs to the ICTT model $\eta$ of 99.9\% is obtained.

\subsection{Impact of Reclosing on Permanent Faults}
The pilot protection based on cosine similarity \cite{cosine} or Pearson coefficient \cite{pearson18} can malfunction when the circuit breaker recloses with a permanent fault.  
To validate the proposed scheme in such scenarios, the grid side breaker allows a $lg$ fault at location 5 with FIT of 11.0s to persist for 2 cycles before opening, then closing after 5 cycles. 756 such permanent faults with auto-reclosing are obtained by varying $v$ (9 and 11m/s), $R_f$ (0.01, 0.1, 1, and 10$\Omega$), and 21 FIT. The proposed scheme identified these faults from non-fault transients without any errors.
\subsection{Influence of Stressed Conditions}
 Distance relays may fail to distinguish stress conditions like power swing, voltage instability, and load encroachment from symmetrical faults and mal-operate. Further, with 3-phase faults during power swings, the power swing blocking may fail to unblock the relay operation and misoperate. The 9-bus test system is used to simulate a variety of typical events, including power swings, three-phase faults during power swings, voltage instability, and load encroachment to validate the scheme. 

\subsubsection{Power Swings}
3-phase to $gnd$ faults are simulated at 11s at the middle of line 4-5 (100 km length) through a $R_f$ of 0.01$\Omega$ while $v$ is maintained at 8, 9, 11, and 22m/s and  the faults are cleared after 3 cycles. As a result, the system experiences 360 cases of stable power swings which are monitored by relay DR2$'$9 at bus-2$'$.

\subsubsection{Faults during Power Swings}
3-phase to $gnd$ faults are simulated at 11s in line 4-5 through a $R_f$ of 0.01$\Omega$ with $v$ at 8, 9, 11, and 22m/s and then cleared after 3 cycles which cause stable symmetrical power swings. 10 different phase \& $gnd$ faults are then simulated at locations 2, 5, and 6 through $R_f$= 0.01$\Omega$, 0.1$\Omega$, 1$\Omega$ and 10$\Omega$  at 11 FIT. Thus, 1320 cases of faults during symmetrical swings are obtained.

\subsubsection{Voltage Instability}
Voltage instability scenarios are created by increasing the reactive power demand at bus-9 in successive steps of 0.00079s beginning at 11.0s. The voltage profiles at this bus decrease progressively resulting in 21 voltage instability cases.

\subsubsection{Load Encroachment}
The load encroachment scenario is developed by increasing the
active power demand of the load connected to bus-9. The load demand is increased up to 150\% of the actual rating  in successive steps of 0.00079s beginning at 11.0s. Consequently, 84 load encroachments or heavily loaded conditions are created.

$\eta$ of 98.5\% is obtained on 1320 faults during power swings and 465 cases of voltage instability, load encroachment, and power swings. Further, $\eta$ of 98.0\% for faults and faults during power swings (11520+1320)  and voltage instability, load encroachment, power swings, capacitor switching, and load switching (465+2400 original cases; 12840 after SMOTE) reaffirm the validity of the findings.


\subsection{Impact of High Impedance Faults}
High impedance faults (HIFs) are difficult to detect using traditional distance or overcurrent relays because of relay sensitivity issues with extremely low-level fault currents and/or relay configuration restrictions. Two anti-parallel DC sources with two diodes and variable resistances are used to model the HIFs \cite{hif,hif2}. The $lg$ fault in phase-a with $R_f$ between 50$\Omega$ to 300$\Omega$ varied randomly in an interval of 2ms are simulated to obtain 555 HIFs faults at locations 4, 5, 6, 7, and 8 with $v$ (9, 11 and 22m/s) for 37 FIT. The proposed scheme identified the HIFs from other transients with 98.3\% $\eta$.

\begin{table*}
\centering
\scriptsize
\renewcommand{\arraystretch}{1.1}
\setlength{\tabcolsep}{2pt}
\caption{Comparison with Recently Reported Methods}
\vspace{-1mm}
\label{ComparisonTable}
\begin{tabular}{ccccccccccccccccccccc} 
\hline
\multirow{2}{*}{\begin{tabular}[c]{@{}c@{}}Year \\\&\\reference\end{tabular}} & \multirow{2}{*}{Method}                                                                                               & \multirow{2}{*}{\begin{tabular}[c]{@{}c@{}}System\\freq. \& \\sampling\\freq.(kHz)\end{tabular}} & \multicolumn{13}{c}{Different Scenarios Studied}                                                                                                                                                                                                                                                                                                                                                                                                                                                                                                                                                                                                                                                                                                  & \multirow{2}{*}{\begin{tabular}[c]{@{}c@{}}Time\\(ms)\end{tabular}} & \multicolumn{4}{c}{$\eta$(\%)}  \\ 
\cline{4-16}\cline{18-21}
                                                                         &                                                                                                                       &                                                                                               & HIF & noise & \begin{tabular}[c]{@{}c@{}}FACTS\\devices \\used\end{tabular}                          & \begin{tabular}[c]{@{}c@{}}sync.\\errors\end{tabular} & \begin{tabular}[c]{@{}c@{}}double\\/single\\-end\end{tabular} & \begin{tabular}[c]{@{}c@{}}reclose\\perm.\\faults\end{tabular} & \begin{tabular}[c]{@{}c@{}}cross-ctry.\\ evolving\\faults\end{tabular} & \begin{tabular}[c]{@{}c@{}}weak \\grid\end{tabular} & \begin{tabular}[c]{@{}c@{}}stress \\cond.\end{tabular} & \begin{tabular}[c]{@{}c@{}}double\\circuit\\line\end{tabular} & \begin{tabular}[c]{@{}c@{}}CT \\sat.\end{tabular} & \begin{tabular}[c]{@{}c@{}}$\Delta$ (samp.\\freq.) \&\\window\end{tabular} & \begin{tabular}[c]{@{}c@{}}cap.\& \\load\\switch\end{tabular} &                                                                     & FD   & FRI   & FC    & FL        \\ 
\hline
\rowcolor[rgb]{0.95,0.95,0.95} 2022\cite{cosine22}                                                                  & cosine similarity                                                                                                     & 50, 1.2                                                                                       & \xmark & \xmark   & \xmark                                                                                    & \xmark                                                   & double                                                       & \cmark                                                            & \cmark                                                                    & \xmark                                                 & \xmark                                                    & \xmark                                                           & \xmark                                               & \xmark                                                                    & \xmark                                                           & 14                                                                  & 100  & \xmark  & \xmark   & \xmark       \\

 2021 \cite{spearman21}                             & spearman correlation                                                                                                  & 50, 5                                                                                          & \cmark & \cmark   & \xmark                                                                                    & \cmark                                                   & double                                                       & \cmark                                                            & \xmark                                                                    & \cmark                                                 & \xmark                                                    & \xmark                                                           & \cmark                                               & \xmark                                                                    & \xmark                                                        &                             20                                        & 100  & \xmark  & \xmark   & \xmark       \\
\rowcolor[rgb]{0.95,0.95,0.95} 2022 \cite{saber22}                           & signed correlation                                                                                                    & 60, 1.2                                                                                        & \xmark & \cmark   & \xmark                                                                                    & \cmark                                                   & double                                                       & \xmark                                                            & \xmark                                                                    & \xmark                                                 & \xmark                                                    & \xmark                                                           & \xmark                                               & \xmark                                                                    & \cmark                                                           & \textasciitilde{}16                                                 & 100  & \xmark  & \xmark   & \xmark       \\
2018 \cite{pearson18}                                                               & pearson correlation                                                                                                   & 50, 1                                                                                          & \xmark & \cmark   & \xmark                                                                                    & \xmark                                                   & double                                                       & \xmark                                                            & \xmark                                                                    & \xmark                                                 & \xmark                                                    & \xmark                                                           & \xmark                                               & \cmark                                                                    & \xmark                                                           & 10                                                                  & 100  & \xmark  & \xmark   & \xmark       \\
\rowcolor[rgb]{0.95,0.95,0.95} 2021 \cite{sauviktcsc21}                          &  EMD, RF & 50, 1                                                                                          & \xmark & \cmark   & TCSC                                                                                   & \xmark                                                   & double                                                       & \xmark                                                            & \xmark                                                                    & \xmark                                                 & \xmark                                                    & \xmark                                                           & \cmark                                               & \xmark                                                                    & \cmark                                                           & 8                                                                   & 100  & \xmark  & 99.95 & 99.12     \\
2022 \cite{sauvikstatcom22}                                                            & dual-time tranform,DT                                                                                                & 50,(1,10)                                                                                     & \xmark & \cmark   & Statcom                                                                                & \xmark                                                   & double                                                       & \xmark                                                            & \xmark                                                                    & \xmark                                                 & \xmark                                                    & \xmark                                                           & \cmark                                               & \xmark                                                                    & \cmark                                                           & 8                                                                   & 100  & 100  & 99.96 & 99.18     \\

\rowcolor[rgb]{0.902,0.969,0.976} proposed                               & \begin{tabular}[c]{@{}>{\cellcolor[rgb]{0.902,0.969,0.976}}c@{}}autoregressive coeff.,\\fuzzy, ICTT\end{tabular}  & 60, 7.68                                                                                       & \cmark & \cmark   & \begin{tabular}[c]{@{}>{\cellcolor[rgb]{0.902,0.969,0.976}}c@{}}PST\\TCSC\end{tabular} & \cmark                                                   & both                                                         & \cmark                                                            & \cmark                                                                    & \cmark                                                 & \cmark                                                    & \cmark                                                           & \cmark                                               & \cmark                                                                    & \cmark                                                           & \textasciitilde10.6                                                                 & 99.6 & 99.0 & 94.0    & 96.2     \\
\hline
\multicolumn{21}{l}{~FD= fault detection, FRI=fault region identification, FC=fault classification, FL=fault location} 
\end{tabular}
\vspace{-1mm}
\end{table*}

\begin{figure*}
\vspace{0.4cm}
\centering
\captionsetup{textfont=normal}
\includegraphics[width=7.15 in, height=1.93 in]{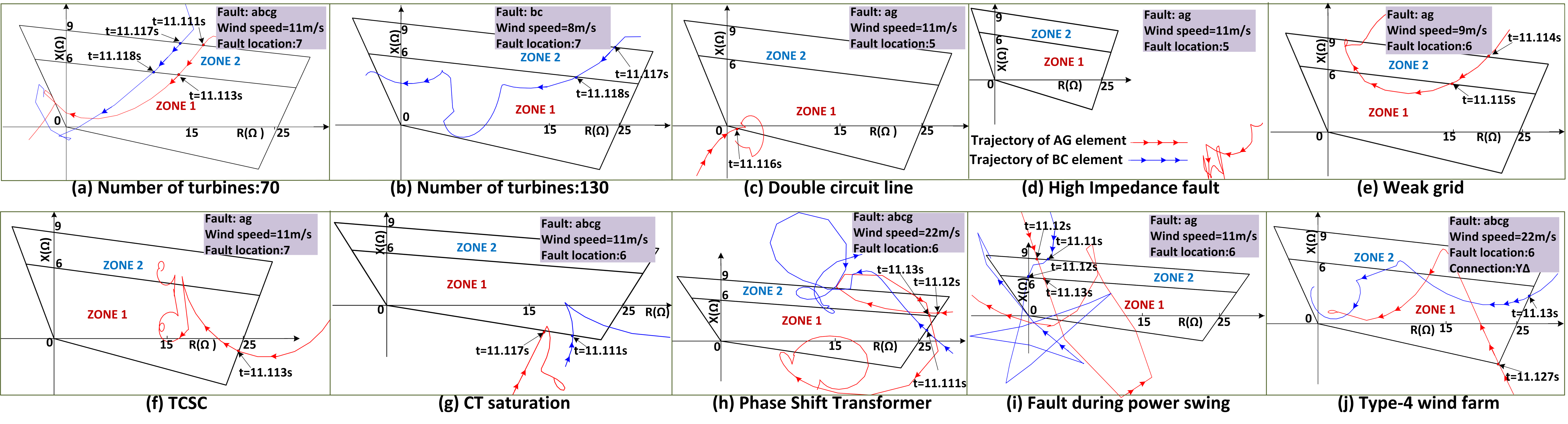}
\vspace{-0.3cm}
\caption{\small{Trajectory of AG and BC elements of the distance relay at $CT_w$ for different fault scenarios --- (a) and (b): change in number of wind turbines, (c): double circuit line fault, (d): high impedance fault, (e): fault with WF connected to weak grid, (f): fault in presence of TCSC, (g): CT saturation, (h): fault in presence of PST, (i): fault during power swing, and (j): fault with type-4 WF.}}
\label{10wf}
\vspace{-2mm}
\end{figure*}

\subsection{Influence of CT Saturation and CCVT Transients}
The CT cores may saturate during severe faults which can adversely affect the performance of protection algorithms. The AR coefficient-based fault detection is immune to faults with CT saturation which is evident from the $\eta$ of 98.8\% obtained on 1760 faults simulated by varying $v$ (8 \& 11m/s), $R_f$ (0.01, 1, 2, \& 10$\Omega$), 10 fault types, and 11 FIT at fault locations 5 \& 6; and 2400 non-faults cases. A higher secondary burden of  20$\Omega$ was chosen so that the CTs get saturated. Since phase currents are considered here for feature extraction the proposed method is not affected by CCVT transients which normally cause overreach for distance relays.

\subsection{Impact of Sampling Rate and Data Size Window}
Data sampling frequency and size of the data window can influence the speed and reliable operation of the protective relay. In this regard, the 3-phase relay currents are sampled at 3.84 kHz and 7.68 kHz. $\eta$ values of 98.6\% and 99.6\%
respectively, shows that the proposed scheme is unaffected by the number of samples used to train the network. Further, the effect of the data window was tested with different window sizes using 0.5-cycle, 1-cycle, 2-cycle, and 3-cycle data. $\eta$ of 99.6\%, 99.6\%, 99.5\%, and 99.5\% respectively demonstrate the robustness of the system to variation in window size. 

\subsection{Impact of Synchronization Errors}
{While evaluating the proposed algorithm for double-end measurement,} when 3-phase currents at the near and far ends are considered, the angle difference between the two ends due to synchronization errors plus the CT angle error results in a time delay of about 1ms \cite{spearman21}. To test the influence of synchronization delay, current waveforms on the grid side obtained at 7.68kHz are delayed by 1ms. $\eta$ of 99.6\% and 99.4\% obtained with and without synchronization delay shows that synchronization errors have negligible impact on the fault detection scheme.
\subsection{Ability to overcome Noise}
Current waveforms are susceptible to noise due to electromagnetic interference.
The noise level of a 220 kV TL is typically 25dB \cite{spearman21}. The anti-noise ability of the suggested method is tested using AR coefficients from 3-phase currents with a signal-to-noise ratio (SNR) of 20dB to 40dB. Gaussian white noise is added to the 3-phase currents recorded at $CT_g$ and $CT_w$. It is seen that in the worst case with 20dB SNR, the overall $\eta$ decreases from 99.4\% to 77\% for single-end measurement ($CT_w$) and 99.6\% to 82\% for double-end measurement ($CT_g$ and $CT_w$). The dependability of the system is unaffected, but the security suffers (Table \ref{noise}). {The adoption of FFT coefficients enhances the security. It is discovered that the absolute values of FFT coefficients (0, 1, and 2) are more reliable in the presence of noise (table \ref{noise}).}

\begin{table}
\vspace{5mm}
\centering
\renewcommand{\arraystretch}{1.1}
\setlength{\tabcolsep}{1.5pt}
\footnotesize
\caption{Effect of Noise}
\label{noise}

\begin{tabular}{cccccc} 
\hline
\multirow{2}{*}{Fault/disturbances}         & \multirow{2}{*}{SNR(dB)} &  \multicolumn{2}{c}{AR coeff. $\eta$(\%)} & \multicolumn{2}{c}{FFT coeff. $\eta$(\%) }                                                  \\ 
\hhline{~~----}
                                            &                           & single-end                                  & double-end & single-end   & double-end    \\ 
\hline
\multirow{4}{*}{Faults}                     & $\infty $                 & 99.8                                        & 99.9                                                            &  99.3 &  99.4 \\
                                            & 40                        & 94.1                                        & 99.2                                                            & 99.2 & 99.4  \\
                                            & 30                        & 92.1                                        & 98.9                                                            & 99.0 & 99.2  \\
                                            & 20                        & 89.6                                        & 91.8                                                            & 98.6 & 99.0  \\ 
\hline
\multirow{4}{*}{\begin{tabular}[c]{@{}c@{}}Load \&\\~Capacitor \\Switching\end{tabular}} & $\infty $                 & 97.3                                        & 97.3                                                            & 99.3 & 99.4  \\
                                            & 40                        & 65.1                                        & 96.1                                                            & 99.1 & 99.2  \\
                                            & 30                        & 45.2                                        & 90.1                                                            & 99.0 &  99.2 \\
                                            & 20                        & 22.4                                        & 34                                                              & 98.8 & 98.9  \\
\hline
\end{tabular}
\end{table}

\subsection{Performance Evaluation with Semi-supervised Learning}
Many power system events are recorded unlabeled, and as supervised ML algorithms can only handle labeled data, the inclusion of such unlabeled data would provide a challenge and cause fault misinterpretation.
Sometimes unsupervised pre-training may improve performance over fully supervised learning.
In order to assess the performance of the proposed scheme, the fault detection model is evaluated with semi-supervised learning using self-training as in \cite{tstmvp}.
The 11520 faults and 2400 non-fault events are pre-trained by removing the target class labels from the data and then this architecture having the pre-trained model weights is trained again with 5\%, 25\% and 50\% labeled datasets. $\eta$ of 90.3\%, 96.6\% and 98\% respectively show that fully supervised ICTT model ($\eta$ of 99.6\%)  proposed in this study performs better.

{Incorrect operation  of distance relay at}  $CT_w$ for faults at different locations which are outside zone 1 of DR2$'$9 for different wind speeds, fault types, and transformer connections with FIT of 11.0s and $R_f$=0.01$\Omega$ for some of the above mentioned scenarios are shown in (Fig.\ref{10wf}(a) and \ref{10wf}(b)--change in number of wind turbines, Fig.\ref{10wf}(c)--double circuit line fault,  Fig.\ref{10wf}(e)--weak grid fault, Fig.\ref{10wf}(f)--fault in presence of TCSC, Fig.\ref{10wf}(g)--fault during CT saturation, Fig.\ref{10wf}(h)--fault in presence of PST, Fig.\ref{10wf}(i)--fault during power swing, and Fig.\ref{10wf}(j)--fault in presence of type-4 WF). Misoperation of the distance relay for a high-impedance fault inside zone 1 is shown in Fig.\ref{10wf}(d).

\section{Comparative Analysis}
{In order to assess the thoroughness of the proposed approach, Table \ref{ComparisonTable} compares many recent publications on criteria, focusing primarily on the influence of various scenarios taken into consideration on the methods proposed in those articles.} It is observed that while the approaches have shown positive outcomes under some specific conditions, performance analysis under many crucial circumstances is missing. Further, the above articles mention the operation time of the proposed algorithms. As per IEEE standard\cite{standard2}, the communication channels between the two terminals for double-ended protection can have a time delay of no more than 6.87ms, which includes delays from the relay interface, fiber optic propagation, and the substation multiplexer.  {Considering a 0.5 cycle (64 samples) the proposed method can trip all faults within $\sim$10.6ms for single-end and within $\sim$17.5ms for double-end measurements (with communication time delay) which is comparable to the aforementioned methods. {Intel Core i5-9500 CPU @ 3.0 GHz and 16 GB RAM is used for simulations, data processing, and evaluation of different algorithms.  }
\section{CONCLUSIONS}
A protection system should meet the five criteria of dependability, security, selectivity, robustness, and speed.
However, the impedance measured by distance relays may be incorrect in cases when the current and voltage frequencies are different. In such events, the dependability for in-zone faults, and security for out-of-zone faults and other non-fault transients is jeopardized.
 The proposed AR coefficient-based intelligent protection scheme is dependable for internal faults, faults during power swings, cross-country faults, evolving faults, auto-reclosing on permanent faults, faults in double circuit lines, and faults with CT saturation; and sensitive to low current for high impedance faults. It is secure to capacitor and load switching, voltage instability, power swings, and load encroachments.  
It is selective and locates the faults accurately. It is robust to changes in the capacity of the WF, type of WF, grid strength, presence of FACTS devices,  single- and double-end measurements, sampling rate, data window size, noise in measurements, and synchronization error.
The AR-based intelligent scheme is tested successfully for a wide variation in fault parameters and operating scenarios in the IEEE 9 and IEEE 39-bus systems.
It is observed that the AR-based adaptive fuzzy fault decision-making system supervised by the AR-trained InceptionTime network is reliable, accurate, and fast. It offers complete protection for transmission lines connected to large wind farms.
\bibliographystyle{IEEEtran}
\bibliography{REFERENCES}

\begin{IEEEbiography}[{\includegraphics[width=1in,height=1.27in,clip,keepaspectratio]{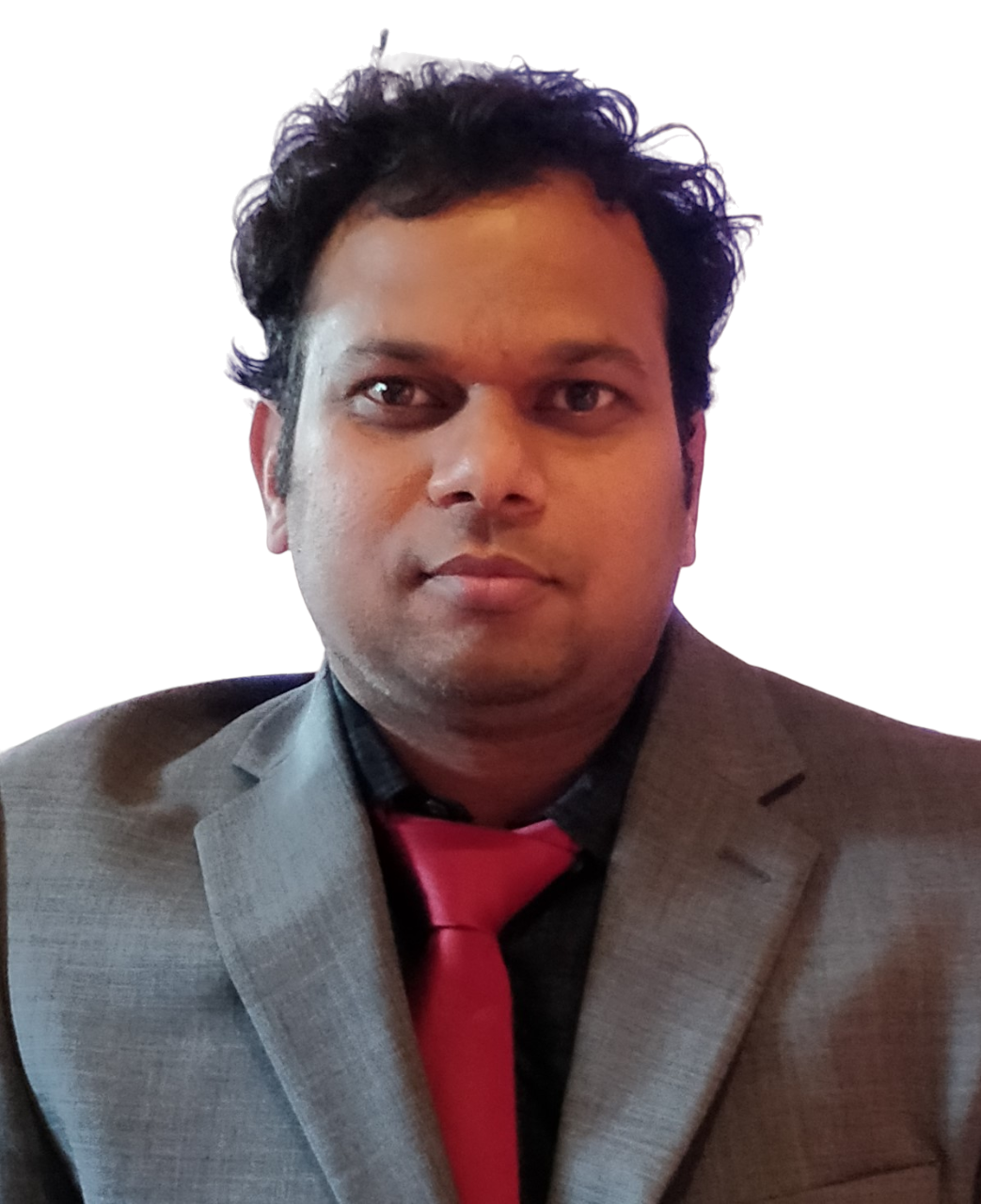}}]{Pallav Kumar Bera} received the bachelor’s degree in electrical engineering from the Haldia Institute of Technology, West Bengal, India, in 2011, the master’s degree in systems and control from the Indian Institute of Technology, Roorkee, India, in 2014, and the Ph.D. degree from the Syracuse University, New York, USA, in 2021. He is currently working as an Assistant Professor at the School of Engineering and Applied Sciences, Western Kentucky University, Bowling Green, Kentucky, USA. His research interests include power system protection and transient analysis. \end{IEEEbiography}
\balance
\begin{IEEEbiography}[{\includegraphics[width=1in,height=1.25in,clip,keepaspectratio]{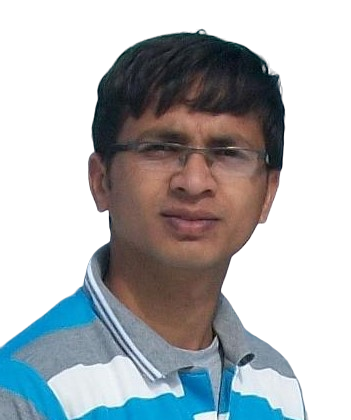}}]{Vajendra Kumar} received the bachelor’s degree in electrical engineering from the Rungta College of Engineering and Technology, Bhilai, Chhattisgarh, India, in 2012, and the master’s degree in instrumentation and signal processing from the Indian Institute of Technology Roorkee, India, in 2014. He currently works as an independent researcher. His research interests include power system protection and transient analysis.
\end{IEEEbiography}

\begin{IEEEbiography}[{\includegraphics[width=1in,height=1.25in,clip,keepaspectratio]{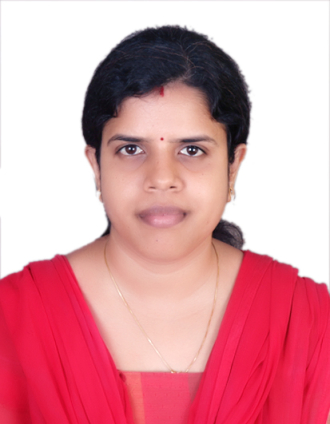}}]{Samita Rani Pani} received the bachelor’s degree in electrical engineering from the Indira Gandhi Institute of Technology, Sarang, India, in 2011, the master’s degree in power system engineering from Veer Surendra Sai University of Technology, Burla, India, in 2014, and currently pursuing the Ph.D. degree with Veer Surendra Sai University of Technology, Burla. She is also working as an Assistant Professor at the School of Electrical Engineering, KIIT deemed to be University, Bhubaneswar, Odisha, India. Her research interests include power system protection and vulnerability.
\end{IEEEbiography}

 \begin{IEEEbiography}[{\includegraphics[width=1in,height=1.25in,clip,keepaspectratio]{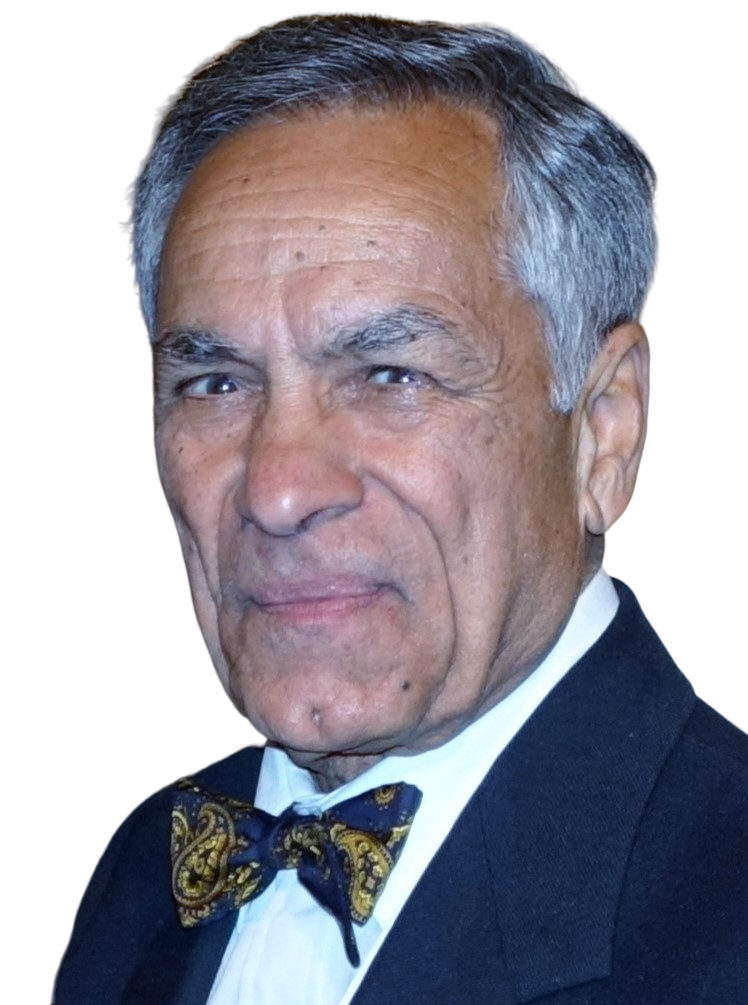}}] {Om P. Malik} has done pioneering work in the development of controllers for application in electric power systems and wind power generation over the past over 50 years. After extensive testing, the adaptive controllers developed by his group are now employed on large generating units. His other interests include digital protection, control of renewable power generation and micro-grids, and AI applications in power system control.

He has published over 800 papers including over 410 papers in international Journals and is the coauthor of four books: (1) Electric Distribution Systems, (2) Power System Stability, (3) Power Grids with Renewable Energy, Power System Stability and Cointrol.

Professor Malik graduated in 1952 from Delhi Polytechnic. After working for nine years in electric utilities in India, he obtained a master’s degree from Roorkee University in 1962, a Ph.D. from London University and a DIC from the Imperial College, London in 1965.

He was teaching and doing research in Canada from 1966 to 1997 and continues to do research as Professor Emeritus at the University of Calgary. Over 100, including 54 Ph.D., students have graduated under his supervision.

Professor Malik is a Life Fellow of IEEE, and a Fellow of IET, the Engineering Institute of Canada, Canadian Academy of Engineering, Engineers Canada and World Innovation Foundation. He is a registered Professional Engineer in the Provinces of Alberta and Ontario, Canada, and has received many awards. He was Director, IEEE Region 7 and President, IEEE Canada during 2010-11 and President, Engineering Institute of Canada, 2014-2016.
\end{IEEEbiography}

\end{document}